\documentclass[iop,apj,revtex4]{emulateapj}
\usepackage{graphicx,url}

\shorttitle{Carbon in dSphs}
\shortauthors{Kirby et al.}
\slugcomment{Accepted to ApJ on 2015 Jan 27}
\newcommand{\ngctot}{154}

\newcommand{\ndsphgctot}{745}
\newcommand{\ntotch}{552}
\newcommand{\ndsphch}{398}
\newcommand{\ndsphul}{182}
\newcommand{\ncstar}{11}

\newcommand{\ncstarfor}{2}
\newcommand{\ncstarumi}{4}
\newcommand{\ncstardra}{5}
\newcommand{\nscl}{181}
\newcommand{\nfor}{45}
\newcommand{\numi}{41}
\newcommand{\ndra}{131}

\newcommand{\cfebump}{0.12}
\newcommand{\cfeint}{1.42}
\newcommand{\cfeinterr}{0.06}
\newcommand{\cfeslope}{0.82}
\newcommand{\cfeslopeerr}{0.02}
\newcommand{\deltaa}{1.6}
\newcommand{\deltab}{1.5}
\newcommand{\deltac}{2.1}
\newcommand{\gccrichname}{47866}
\newcommand{\gccrichcfe}{+0.88}
\newcommand{\gccrichcfeerr}{0.10}

\newcommand{\residscl}{1.5}
\newcommand{\residfor}{1.2}
\newcommand{\residumi}{1.5}
\newcommand{\residdra}{1.6}

\newcommand{\fehb}{-3.0}
\newcommand{\ntb}{2}
\newcommand{\ncb}{1}
\newcommand{\cempfracb}{50}
\newcommand{\cempfracerrb}{50}
\newcommand{\fehc}{-2.5}

\newcommand{\ncc}{2}
\newcommand{\cempfracc}{7.1}
\newcommand{\cempfracerrc}{5.1}

\newcommand{\fehf}{-1.0}

\newcommand{\ncf}{5}
\newcommand{\cempfracf}{1.3}
\newcommand{\cempfracerrf}{0.6}

\newcommand{\ncemp}{6}
\newcommand{\ntot}{398}
\newcommand{\ncemppla}{3}
\newcommand{\cempplafeha}{-2.51}
\newcommand{\cempplafeherra}{0.26}
\newcommand{\cempplafehb}{-2.07}
\newcommand{\cempplafeherrb}{0.11}
\newcommand{\cempplafehc}{-3.22}
\newcommand{\cempplafeherrc}{0.20}
\newcommand{\cfecorrempa}{-0.39}
\newcommand{\cfeerrempa}{0.37}
\newcommand{\cfecorrempb}{+0.76}
\newcommand{\cfeerrempb}{0.13}
\newcommand{\shecstarcfe}{-0.52}
\newcommand{\shecstarcfeerr}{0.11}

\newcommand{\feh}{{\rm [Fe/H]}}
\newcommand{\cfe}{{\rm [C/Fe]}}
\newcommand{\cfecorr}{{\rm [C/Fe]}_{\rm corr}}

\begin{document}

\title{Carbon in Red Giants in Globular Clusters and Dwarf Spheroidal Galaxies\altaffilmark{*}}

\author{Evan~N.~Kirby\altaffilmark{1},
  Michelle~Guo\altaffilmark{2,3},
  Andrew~J.~Zhang\altaffilmark{4},
  Michelle~Deng\altaffilmark{5},
  Judith~G.~Cohen\altaffilmark{1}, 
  Puragra~Guhathakurta\altaffilmark{6},
  Matthew~D.~Shetrone\altaffilmark{7},
  Young~Sun~Lee\altaffilmark{8},
  Luca~Rizzi\altaffilmark{9}}

\altaffiltext{*}{The data presented herein were obtained at the
  W.~M.~Keck Observatory, which is operated as a scientific
  partnership among the California Institute of Technology, the
  University of California and the National Aeronautics and Space
  Administration. The Observatory was made possible by the generous
  financial support of the W.~M.~Keck Foundation.}
\altaffiltext{1}{California Institute of Technology, 1200 E.\ California Blvd., MC 249-17, Pasadena, CA 91125, USA}
\altaffiltext{2}{Irvington High School, 41800 Blacow Rd., Fremont, CA 94538, USA}
\altaffiltext{3}{Stanford University, 450 Serra Mall, Stanford, CA 94305, USA}
\altaffiltext{4}{The Harker School, 500 Saratoga Ave., San Jose, CA 95129, USA}
\altaffiltext{5}{Harvard University, Massachusetts Hall, Cambridge, MA 02138, USA}
\altaffiltext{6}{UCO/Lick Observatory and Department of Astronomy and Astrophysics, University of California, 1156 High St., Santa Cruz, CA 95064, USA}
\altaffiltext{7}{McDonald Observatory, University of Texas at Austin, HC75 Box 1337-MCD, Fort Davis, TX 79734, USA}
\altaffiltext{8}{Chungnam National University, Department of Astronomy and Space Science, Daejeon 305-764, Republic of Korea}
\altaffiltext{9}{Keck Observatory, 65-1120 Mamalahoa Hwy, Kamuela, HI 96743, USA}



\begin{abstract}

We present carbon abundances of red giants in Milky Way globular
clusters and dwarf spheroidal galaxies (dSphs).  Our sample includes
measurements of carbon abundances for \ngctot\ giants in the clusters
NGC~2419, M68, and M15 and \ntot\ giants in the dSphs Sculptor,
Fornax, Ursa Minor, and Draco.  This sample doubles the number of dSph
stars with measurements of [C/Fe].  The [C/Fe] ratio in the clusters
decreases with increasing luminosity above $\log (L/L_{\sun}) \simeq
1.6$, which can be explained by deep mixing in evolved giants.  The
same decrease is observed in dSphs, but the initial [C/Fe] of the dSph
giants is not uniform.  Stars in dSphs at lower metallicities have
larger [C/Fe] ratios.  We hypothesize that [C/Fe] (corrected to the
initial carbon abundance) declines with increasing [Fe/H] due to the
metallicity dependence of the carbon yield of asymptotic giant branch
stars and due to the increasing importance of Type~Ia supernovae at
higher metallicities.  We also identified \ncstar\ very carbon-rich
giants (8 previously known) in three dSphs.  However, our selection
biases preclude a detailed comparison to the carbon-enhanced fraction
of the Milky Way stellar halo.  Nonetheless, the stars with $\cfe <
+1$ in dSphs follow a different [C/Fe] track with [Fe/H] than the halo
stars.  Specifically, [C/Fe] in dSphs begins to decline at lower
[Fe/H] than in the halo.  The difference in the metallicity of the
[C/Fe] ``knee'' adds to the evidence from [$\alpha$/Fe] distributions
that the progenitors of the halo had a shorter timescale for chemical
enrichment than the surviving dSphs.

\end{abstract}


\section{Introduction}
\label{sec:intro}

Variations of elemental abundances within a stellar system are
evidence for physical processes both internal and external to
individual stars.  Chemical evolution on a galactic scale leads to a
dispersion in metallicity.  As progressive generations of supernovae
enrich the interstellar medium of a galaxy, stars that form later have
higher metallicity than their older predecessors \citep{tin80}.  On
the other hand, variations in certain light elements can arise from
internal stellar evolution.  Gravitational settling and radiative
levitation \citep{beh99} deplete and enhance, respectively, the
abundance of an element in the photosphere of a star relative to the
rest of the star.  Nuclear burning at the base of a convective zone
that reaches the photosphere can also decrease or increase the
observable abundances of light elements, like Li \citep{lin11}, C, N,
O, and even $s$-process elements, like Ba and Rb \citep{lia00,kar12}.
This process is called astration when nuclear burning leads to the
destruction of an element.

Carbon exhibits abundance variations within a stellar population due
to astration and galactic chemical evolution.  As a star ascends the
red giant branch (RGB), it encounters episodes of mixing that bring
material processed through the CNO cycle to the photosphere.  Within
2--3 magnitudes of the tip of the RGB, C and O are depleted and N is
enhanced relative to stars on the main sequence or the lower RGB\@.
Proposed mechanisms for mixing deep CNO-processed material into the
upper atmosphere include Rayleigh-Taylor instability \citep{egg06} or,
more likely, thermohaline mixing \citep{cha07}.

Globular clusters (GCs) are excellent sites to study the evolution of
carbon.  It is relatively easy to test whether variations in carbon
abundances within a cluster are primordial or modified by stellar
evolutionary processes during the lifetime of the cluster because GCs
are essentially single-age, mono-metallic populations at a uniform
distance.  It is not necessary to control for age or metallicity.
Carbon should not be astrated in metal-poor stars on the main sequence
and subgiant branch, which do not have convective envelopes deep
enough to dredge up CNO-processed material.  However, GCs do show
primordial abundance variations.  \citet{coh99b} and \citet{bri02}
found variations in [C/Fe] on the main sequences of M71 and M13,
respectively.  Because the photospheres of main sequence stars are not
affected by convection that can reach CNO-processed material, these
carbon variations must be primordial.  Regardless, astration on the
upper RGB causes the most severe variations
\citep[e.g.,][]{sun81,smi96,smi05,smi06}.  \citet{mar08c} showed that
the rate of depletion depends on metallicity, as predicted by
thermohaline mixing \citep{cha07}.

Stars can also become very carbon-rich.  ``Carbon stars'' \citep[see
  the review by][]{wal98} display extremely strong C$_2$, CH, and/or
CN bands, and they can be significantly redder than the RGB in optical
colors \citep[e.g.,][]{bon69,mou79}.  Carbon can form in
helium-burning asymptotic giant branch (AGB) stars through the
triple-$\alpha$ process.  It can be brought to the surface of an AGB
star in the third dredge-up \citep{ibe75}.  This process is most
efficient in stars around 2--2.5~$M_{\sun}$ \citep{mou03}, which
explains why most ancient GCs---where the highest-mass stars are less
than 1~$M_{\sun}$---lack C stars \citep{coh78}.  C stars also require
at least 0.3~Gyr to form \citep{mou02}, which makes them good
indicators of intermediate-age populations.

Dwarf and RGB stars can acquire carbon through binary mass transfer
with a C star \citep[see][]{luc05,coh06,sta14}.  C-rich RGB stars are
rarely seen to have a photometric companion because the AGB donor has
almost always evolved into a faint white dwarf long ago.  Instead,
binarity of C-rich RGB stars is observed spectroscopically through
variability in radial velocity \citep{mcc90}.  The recipient of carbon
through mass transfer can exist in any evolutionary stage, such as the
main sequence \citep{dah77}.

\citet{aar80} and \citet{aar82} used the existence of carbon stars in
dwarf spheroidal galaxies (dSphs) to argue for the presence of
intermediate-age populations.  These stars were selected for
spectroscopy from their infrared colors, and their spectra showed very
strong C$_2$ bands.  However, carbon enhancements need not be so large
that they affect broadband colors.  For example, \citet{nor10}
discovered a C-rich ($\cfe = 2.3$) red giant in the ultra-faint galaxy
Segue~1.  This star (Segue~1--7) could only have been found
spectroscopically because its broadband colors are consistent with
carbon-normal RGB stars.

The statistics of carbon abundances in dwarf galaxies are less
complete than for GCs or the Milky Way (MW)\@.  Although \citet{aar80}
and their contemporaries discovered carbon stars in dwarf galaxies,
they did not quantify the carbon enhancement.  \citet{kin81} published
in conference proceedings the first carbon abundances from
low-resolution spectroscopy in a dSph, specifically Draco.
\citet{smi84} followed up with CH and CN indices in Draco, and
\citet{bel85} published more low-resolution [C/H] abundances in Draco
and Ursa Minor.  \citet{she98b} later measured low-resolution [C/Fe]
abundances for two stars in Sculptor, finding one to be mildly
carbon-rich.  \citet{ful04} measured the first carbon abundance from
high-resolution spectroscopy in a dSph, specifically Draco.
\citet{abi08} measured the $^{12}$C/$^{13}$C isotopic ratio and the
C/O ratio---but not [C/H]---in two carbon stars in Carina.
\citet{coh09,coh10} measured carbon abundances in 18 stars in Draco
and Ursa Minor from the G band of CH, and \citet{sku15} measured C
abundances for about 80 Sculptor giants from a near-infrared CN band.
Recently, ultra-faint galaxies have received a lot of attention,
including carbon abundance measurements in Segue~1
\citep{nor10,fre14}, Ursa Major~II and Coma Berenices \citep{fre10b},
Bo{\" o}tes~I \citep{nor10,lai11}, and Bo{\" o}tes~II \citep{koc14}.
Interestingly, all of these measurements are consistent with GC
abundances on the upper RGB except one star in Sculptor \citep{sku15}
and four stars in Segue~1, including its three most metal-poor stars
\citep{nor10}.  The tendency of carbon-rich stars to appear among the
metal-poor population led \citet{she13} to measure carbon abundances
for 35 stars in the metal-poor dSph Draco and \citet{fre10a} and
\citet{sta13} to explicitly target the most metal-poor stars in
Sculptor.  Still, Segue~1 and Sculptor contain the only examples of
dwarf galaxy stars with quantified high carbon abundances.  (As
discussed above, C stars still exist in dSphs.)

The MW halo contains a large number of C-rich stars, and their
frequency increases with decreasing metallicity.  However, the exact
fraction of carbon-enhanced, metal-poor (CEMP) stars as a function of
metallicity is contentious
\citep{coh05,mar05,luc06,fre06,car12,yon13,lee13,pla14}, partly
because the search for low-metallicity stars is fraught with selection
biases \citep[e.g.,][]{sch09}.  Very roughly, the fraction of stars
with $\feh < -2.5$ that are carbon-enhanced is about 20\%.

The MW halo is expected to be at least partly composed of the
disrupted dSphs from past accretion events \citep{sea78}.  One
apparent challenge to this idea was the discrepancy between the
[$\alpha$/Fe] ratio of halo and dSph stars at $\feh \ga -2$
\citep{she98a,she01,she03,ven04}.  However, \citet{rob05} and
\citet{fon06} showed that the discrepancy is in fact an expected
outcome of hierarchical assembly.  The inner halo, where most of the
observed halo stars lie, is composed of more massive satellites that
do not resemble the present-day, surviving dSphs.  At low enough
metallicities, most elemental abundance ratios between dwarf galaxies
and the MW halo agree \citep[e.g.,][]{fre10b}, though the
neutron-capture elements are deficient in ultra-faint dwarf galaxies
compared to the halo, even at very low metallicities
\citep[e.g.,][]{koc13}.

The fact that so few carbon-rich stars have been found in dSphs, even
at low metallicities, might be troublesome for the theory of the
hierarchical assembly of the MW halo.  The fraction of CEMP stars in
Draco, Ursa Minor, and Sculptor is around 10\% or less
\citep{coh09,coh10,sta13,she13,sku15}.  However, this number does not
include C-rich stars for which carbon abundances were not quantified.

In this work, we add to the body of carbon abundance measurements in
dSphs.  We measured carbon abundances of \ndsphch~red giants (8
previously known) in the dSphs Sculptor, Fornax, Ursa Minor, and
Draco.  We also identified \ncstar~giants that are too carbon-rich to
quantify their abundances.  The stars observed here are a subset of
the stars observed by \citet{kir09,kir10}, who measured Mg, Si, Ca,
Ti, and Fe abundances in about 3000 red giants in eight dSphs.


\section{Spectroscopic Observations}
\label{sec:obs}

We obtained Keck/DEIMOS \citep{fab03} spectra of the carbon G band in
red giants in MW GCs and dSphs.  The GCs are NGC~2419, NGC~4590 (M68),
and NGC~7078 (M15).  The dSphs are Sculptor, Fornax, Ursa Minor, and
Draco.  We used the same dSph slitmasks as \citet{kir09,kir10}, who
obtained redder DEIMOS spectra for measuring Mg, Si, Ca, Ti, and Fe
abundances.  We re-observed 3 of 5 slitmasks (198 of 376 member stars)
in Sculptor, 1 of 5 slitmasks (125 of 675 members) in Fornax, 3 of 4
slitmasks (87 of 212 members) in Ursa Minor, and 3 of 8 slitmasks (181
of 298 members) in Draco.

\subsection{Target selection}
\label{sec:selection}

We selected stars for the DEIMOS slitmasks from a variety of
photometric catalogs, which provided astrometry, magnitudes, and
colors of the red giants that we observed.  We used
\citeauthor{ste00}'s (\citeyear{ste00}) catalog for all three GCs.  We
supplemented \citeauthor{ste00}'s list of M68 stars with photometry
from \citet{alc90} and \citet{wal94}.  We used Johnson $V$ and Cousins
$I$ magnitudes for all GC stars.  We corrected the magnitudes for
reddening and extinction based on $E(B-V)$ values of 0.08, 0.05, and
0.10 for NGC~2419, M68, and M15, respectively \citep[][updated
  2010\footnote{\url{http://www.physics.mcmaster.ca/$\sim$harris/mwgc.dat}}]{har96}.

\citet{kir09} described the target selection for Sculptor, and
\citet{kir10} detailed the target selection for Fornax, Ursa Minor,
and Draco.  For Sculptor, we used \citeauthor{wes06}'s
(\citeyear{wes06}) photometric catalog in Washington $M$ and $T_2$
magnitudes.  We used sources from \citeauthor{ste98}'s
(\citeyear{ste98}) catalog in Cousins $B$ and Johnson $V$ magnitudes
for Fornax.  For Ursa Minor, we used \citeauthor{bel02}'s
(\citeyear{bel02}) $VI$ catalog, and for Draco, we used
\citeauthor{seg07}'s (\citeyear{seg07}) $gi$ catalog in the CFHT
MegaCam system.

We placed stars with colors and magnitudes appropriate for red giants
on the slitmasks.  For the GCs, we drew a polygon around the RGB in
the color--magnitude diagram (CMD)\@.  Stars within the polygon were
included in the slitmask.  Brighter stars were preferred over fainter
stars in the cases where slitmask design constraints forced a choice
among multiple red giants.  For the dSphs, we selected stars from the
CMD with the guidance of theoretical isochrones (see \citealt{kir10}
for additional details).  The Sculptor catalog additionally has DDO21
magnitudes, which we used to discriminate between red giants and
foreground dwarf stars (see \citealt{kir09}).

\subsection{Observations}

\begin{deluxetable*}{llcr@{ }c@{ }lcclc}
\tablewidth{0pt}
\tablecolumns{9}
\tablecaption{DEIMOS Observations\label{tab:obs}}
\tablehead{\colhead{System} & \colhead{Slitmask} & \colhead{\# targets} & \multicolumn{3}{c}{Date} & \colhead{Airmass} & \colhead{Seeing} & \colhead{Individual Exposures} & \colhead{Total Exposure Time} \\
\colhead{ } & \colhead{ } & \colhead{ } & \multicolumn{3}{c}{ } & \colhead{ } & \colhead{($''$)} & \colhead{(s)} & \colhead{(s)}}
\startdata
NGC~2419   & n2419b  &     112 & 2012 & Mar & 19 & 1.1  & $0.8$ & $3 \times 900$ & 2700 \\
M68        & n4590a  & \phn 96 & 2014 & Feb & 2  & 1.6  & $0.8$ & $1200 + 937$ & 2137 \\
M15        & 7078d   &     164 & 2011 & Jul & 29 & 1.0  & $1.1$ & $3 \times 600$ & 1800 \\
           & 7078e   &     167 & 2011 & Jul & 30 & 1.0  & $0.9$ & $3 \times 900$ & 2700 \\
Sculptor   & bscl1   & \phn 86 & 2011 & Jul & 31 & 1.7  & $0.8$ & $4 \times 1200 + 900$ & 5700 \\
           & bscl2   &     106 & 2011 & Aug & 6  & 1.8  & $0.7$ & $2 \times 1200 + 2 \times 840$ & 4080 \\
           & bscl6   & \phn 91 & 2011 & Aug & 4  & 1.7  & $0.8$ & $3 \times 1200 + 1260$ & 4860 \\
Fornax     & bfor6   &     169 & 2011 & Aug & 5  & 1.8  & $1.3$ & $2 \times 780$ & 6360 \\
           &         &         & 2011 & Aug & 6  & 1.8  & $0.8$ & $1200$ & \\
           &         &         & 2011 & Aug & 7  & 1.9  & $0.8$ & $3 \times 1200$ & \\
Ursa~Minor & bumi1   &     125 & 2011 & Jul & 29 & 1.5  & $0.7$ & $4 \times 1200 + 600$ & 5400 \\
           & bumi2   &     134 & 2011 & Jul & 31 & 1.7  & $0.8$ & $4 \times 1200$ & 4800 \\
           & bumi3   &     137 & 2011 & Aug & 4  & 1.8  & $0.6$ & $4 \times 1200$ & 4800 \\
Draco      & bdra1   &     135 & 2011 & Jul & 30 & 1.4  & $1.2$ & $5 \times 1200$ & 6000 \\
           & bdra2   &     144 & 2011 & Aug & 7  & 1.3  & $0.7$ & $4 \times 1200$ & 4800 \\
           & bdra3   &     129 & 2011 & Aug & 5  & 1.3  & $1.0$ & $5 \times 1200$ & 6000 \\
\enddata
\end{deluxetable*}

Table~\ref{tab:obs} lists the observation dates and exposure times of
each slitmask.  We used the 900ZD grating with a ruling of 900
lines~mm$^{-1}$ and a blaze wavelength of 5500~\AA\@.  The grating was
tilted such that the blaze wavelength fell on the center of the CCD
mosaic.  Slits were $0.7\arcsec$ wide.  This configuration gives an
approximate wavelength range of 4000--7400~\AA\ at a resolution of
2.1~\AA\ FWHM ($R \sim 2100$ at 4300~\AA)\@.  The exact wavelength
range of each spectrum depends on the location of the slit on the
slitmask.  The starting and ending wavelengths of the spectra vary by
$\sim 500$~\AA\ across the slitmask.

We obtained internal flat field and arc lamp exposures.  We exposed a
quartz lamp three times for 12~s and six times for 45~s, which is
appropriate for the red and blue halves of the CCD mosaic,
respectively.  We also obtained two sets of arc lamp exposures.
First, we obtained simultaneous spectra of Ne, Ar, Kr, and Xe lamps
for 1~s.  This exposure is suitable to calibrate the wavelength
solution for the red halves of the spectra.  Then, we obtained a
single exposure for several lamps.  We controlled the amount of flux
from each lamp by turning off lamps in the middle of the exposure.
The lamps were Hg (1~s), Cd (5~s), Kr (6~s), Ar (12~s), and Zn (12~s).
This exposure is appropriate for the blue halves of the spectra.

All slitmasks were also observed with the 1200G grating, which has a
groove spacing of 1200~mm$^{-1}$ and a blaze wavelength of 7760~\AA\@.
The grating was centered at 7800~\AA\@.  Table 2 of \citet{kir10}
lists all of these observations except for the n2419b and n4590a
slitmasks.  We observed n2419b with the 1200G grating and this
7800~\AA\ setting on 2012 March 19 for 1800~s, and we observed n4590a
on 2011 June 2 for 2400~s.

\subsection{Reductions}

\begin{figure}[t!]
\centering
\includegraphics[width=\columnwidth]{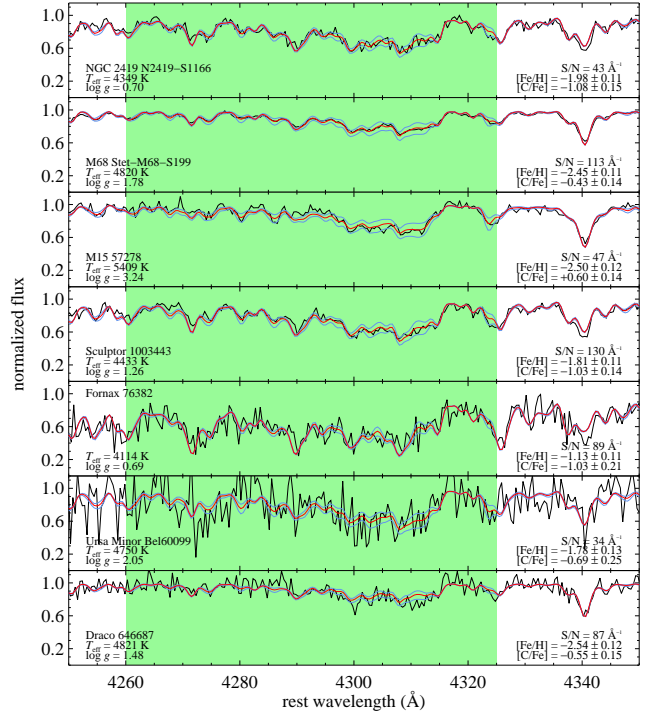}
\caption{Portions of DEIMOS spectra around the G band for one star in
  each of the GCs and dSphs observed.  The black line is the observed
  spectrum, and the red line is the best-fitting synthetic spectrum.
  The fit was performed in the green shaded region.  The pale blue
  lines show synthetic spectra with $\cfe \pm 0.2$ different from the
  best-fitting value.\label{fig:gband}}
\end{figure}

We reduced the raw data with the spec2d
pipeline\footnote{\url{http://www2.keck.hawaii.edu/inst/deimos/pipeline.html}}
\citep{coo12} developed by the DEEP2 survey team \citep{new13}.  This
software is optimized for extended galaxies observed with the 1200G
grating.  We made moderate revisions to the pipeline to make it more
suitable for bluer DEIMOS spectra.  First, we modified the software to
accept separate flat lamp exposures for the red and blue CCDs.
Second, we revised the arc lamp line lists for Hg, Cd, Kr, Ar, and Zn
between 4158~\AA\ and 5496~\AA\@.  We eliminated lines that we could
not see in individual DEIMOS arc lamp exposures.  Third, we changed
the tolerances for automatic identification of arc lines so that the
pipeline would include some of the weaker blue lines.  Fourth, we
changed the way that the two-dimensional spectrum is rectified.  The
curvature of the focal plane and the slit position angle can introduce
a tilt in the spatial direction such that a single wavelength
corresponds to a diagonal line rather than a single row of pixels.
Previously, this tilt was removed by shifting CCD columns by integer
values.  We changed the code to allow non-integer shifts, which
slightly increased the effective spectral resolution and
signal-to-noise ratio (S/N)\@.  Fifth, we implemented a correction for
atmospheric dispersion along the slit, which is important for our
bluer spectra.  Because light at 4000~\AA\ can be displaced along the
slit by a few pixels from light at 7400~\AA\@, we curved the
extraction window to match the shape of the atmospheric dispersion.
Finally, we adopted the optimizations for point sources developed by
\citet{sim07}.

The results of the reductions were wavelength-calibrated,
sky-subtracted, one-dimensional spectra.  Cosmic rays were removed
before the one-dimensional spectra were extracted.  The spec2d code
also calculated and saved variance (error) spectra, which we used to
estimate random measurement uncertainty on [C/Fe]
(Section~\ref{sec:uncertainty}).

Figure~\ref{fig:gband} shows one example spectrum from each stellar
system in Table~\ref{tab:obs}.  The figure shows spectra at a variety
of S/N\@.  Each panel lists the S/N, which is calculated as the median
absolute deviation of the observed spectrum (black) from the
best-fitting synthetic spectrum (red).

\subsection{Carbon Stars}
\label{sec:cstars}

\begin{figure}[t!]
\centering
\includegraphics[width=\columnwidth]{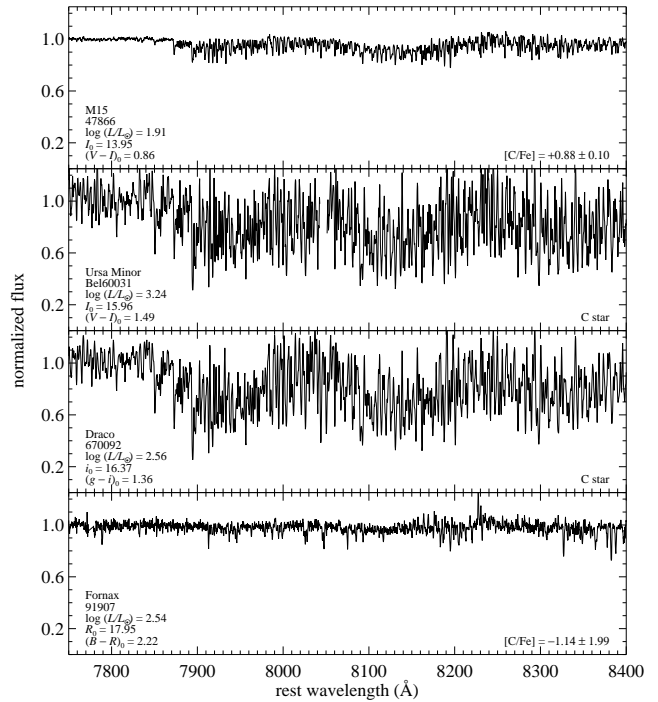}
\caption{The top panel shows a moderately carbon-rich AGB star in M15.
  The CN bandhead is at $\sim 7850$~\AA, and CN absorption is visible
  redward of the bandhead.  The next two panels show CN absorption in
  two of the \ncstar\ carbon stars in the DEIMOS sample.  The bottom
  panel shows a carbon-normal RGB star for
  comparison.\label{fig:cstars}}
\end{figure}

We found \ncstar~examples of carbon stars in our sample.  The red
spectra of these stars are dominated by very strong CN absorption,
which precluded measurement of atmospheric parameters like [Fe/H].
The G bands are also strong enough that we would not have found our
measurements of carbon abundances reliable due to saturation.
Figure~\ref{fig:cstars} shows examples of the near-infrared spectra of
some of these stars.  We discuss carbon stars further in
Sections~\ref{sec:dsphcstars} and \ref{sec:mw}.


\section{Spectroscopic Measurements}
\label{sec:spectroscopy}

Before we could measure carbon abundances from the spectra, we needed
to measure the stars' radial velocities ($v_r$) and atmospheric
parameters.

\subsection{Radial Velocity Measurements}
\label{sec:rv}

We previously measured $v_r$ for all of the spectra from our red
DEIMOS observations \citep{kir09,kir10}.  However, the wavelength
calibration for the blue DEIMOS spectra is not as precise as the red
spectra because of the relative sparsity of blue arc lamp lines.
Rather than apply the known $v_r$ from the red spectra to the blue
spectra, we measured $v_r$ directly from the G band in the blue
spectra.  Although $v_{r,{\rm blue}}$ is not as precise as $v_{r,{\rm
    red}}$ (for example, for measuring galactic velocity dispersions),
$v_{r,{\rm blue}}$ is more suited to shifting the G band into the rest
frame.

We measured $v_{r,{\rm blue}}$ by cross-correlating the observed
spectrum with a synthetic spectrum.  The synthetic spectrum had the
atmospheric parameters previously measured from the red spectra (see
Section~\ref{sec:atm}) and $\cfe = 0.0$.  The spectra were
synthesized according to the procedure described in
Section~\ref{sec:abundance}.  We performed the cross-correlation in
the range 4200--4400~\AA, and we set $v_r$ to be the velocity
corresponding to the cross-correlation peak.  For the remaining
analysis, we shifted the spectrum into the rest frame using this
measurement of $v_r$.

We also evaluated the membership of each star by its radial velocity.
\citet[][their Section~2.6]{kir10} gave the details of the membership
criteria.  In summary, we eliminated stars more than three standard
deviations away from the mean radial velocity.

\subsection{Atmospheric Parameters}
\label{sec:atm}

The measurement of carbon abundances (Section~\ref{sec:abundance})
requires knowledge of effective temperatures ($T_{\rm eff}$), surface
gravities ($\log g$), metallicities ([Fe/H]), and alpha enhancements
([$\alpha$/Fe]).  \citet{kir09,kir10} measured these parameters for
most of the spectra presented here.  They obtained preliminary
estimates of $T_{\rm eff}$ and $\log g$ by comparing the stars' colors
and magnitudes to theoretical isochrones.  They refined $T_{\rm eff}$
and measured [Fe/H] and [$\alpha$/Fe] by minimizing $\chi^2$ between
the observed red DEIMOS spectra and a large grid of synthetic spectra
\citep{kir11}.

We adopted without modification the previously determined atmospheric
parameters.  However, the DEIMOS slitmasks n2419b and n4590a were not
included in the previous catalogs of atmospheric parameters.  We
measured these parameters in the same manner as \citet{kir10}.

\subsection{Grid of Synthetic Spectra}
\label{sec:grid}

We measured [C/Fe] by comparing observed spectra of the G band to a
grid of synthetic spectra spanning 4100--4500~\AA\@.  We computed this
grid in a similar fashion to the way in which \citet*{kir08} computed
their synthetic spectral grid spanning 6300--9100~\AA\ in order to
measure [Fe/H] and [$\alpha$/Fe] from red DEIMOS spectra.  First, we
compiled a line list from publicly available catalogs of atomic and
molecular data.  We refined the line list by comparing it to the
spectra of the Sun and Arcturus.  Then, we computed MOOG synthetic
spectra using ATLAS9 model atmospheres.

\subsubsection{Line List}
\label{sec:list}

\begin{deluxetable}{llcc}
\tablewidth{0pt}
\tablecolumns{4}
\tablecaption{Line List\label{tab:linelist}}
\tablehead{\colhead{Wavelength (\AA)\tablenotemark{a}} & \colhead{Species} & \colhead{EP (eV)} & \colhead{$\log gf$}}
\startdata
4100.005 & Mn$\,${\sc ii} &  8.129 & $-1.184$ \\
4100.013 & NH             &  2.202 & $-3.421$ \\
4100.014 & NH             &  2.387 & $-4.433$ \\
4100.021 & NH             &  2.335 & $-4.289$ \\
4100.022 & $^{13}$CH      &  0.642 & $-4.345$ \\
4100.048 & NH             &  2.824 & $-2.929$ \\
4100.051 & NH             &  2.335 & $-4.321$ \\
4100.055 & NH             &  2.599 & $-2.932$ \\
4100.084 & NH             &  2.202 & $-4.982$ \\
4100.088 & Cr$\,${\sc i}  &  4.535 & $-1.241$ \\
\nodata & \nodata & \nodata & \nodata \\
\enddata
\tablerefs{Atomic data come from VALD \citep{pis95,kup99} and NIST \citep{kra14}.  CH data come from SCAN \citep{jor96}.  NH and CN data come from \citet{kur92}.}
\tablenotetext{a}{Wavelength in air.}
\tablecomments{(This table is available in its entirety in a machine-readable form in the online journal. A portion is shown here for guidance regarding its form and content.)}
\end{deluxetable}

The G band has a high density of CH molecular absorption lines.  At
the low resolution of our DEIMOS spectra, individual CH lines are
indistinguishable from individual atomic and other molecular lines.
Therefore, we needed to model the absorption due to many species, not
just CH\@.  We compiled a list of atomic and molecular transitions in
the range 4100--4500~\AA\@.

We started by including all the neutral and singly ionized atomic
transitions listed in the Vienna Atomic Line
Database\footnote{\url{http://vald.astro.univie.ac.at/$\sim$vald/php/vald.php}}
\citep[VALD,][]{pis95,kup99} in our spectral range of interest.  We
discarded transitions with excitation potentials (EP) greater than
10~eV or with oscillator strengths $\log gf < -5$.

We then compiled a separate atomic line list from the National
Institute of Standards and Technology (NIST)
database\footnote{\url{http://www.nist.gov/pml/data/asd.cfm}}
\citep{kra14}.  This database is very incomplete, but it generally has
more trustworthy, laboratory-measured atomic transition probabilities
than other databases.  We replaced entries in the VALD line list with
NIST entries where available.

Next, we added molecular transitions.  We used the CN and NH line
lists of \citet{kur92}\footnote{\url{http://kurucz.harvard.edu/}} and
the SCAN line
list\footnote{\url{http://www.astro.ku.dk/$\sim$uffegj/}} for CH from
\citet{jor96}.  (\citealt{mas14} published a new CH line list long
after we computed the spectral grid and shortly before we submitted
this article.)  We preserved the isotopic information for each CH
transition so that we could change $^{12}$C/$^{13}$C in the spectral
syntheses.

No line list is perfectly accurate or perfectly complete.  We tested
our own line list against observed, high-resolution spectra of the Sun
and Arcturus \citep{hin00}.  We computed synthetic spectra for both
stars using
MOOG\footnote{\url{http://www.as.utexas.edu/$\sim$chris/moog.html}}
\citep{sne73} with an updated treatment of electron scattering
\citep{sob11}.  For the Sun, we used an ATLAS9 \citep{kur93} model
atmosphere with $T_{\rm eff} = 5798$~K and $\log g = 4.44$.  For
Arcturus, we used \citeauthor{pet93}'s (\citeyear{pet93}) model
atmosphere.  Where atomic lines did not match the observed spectra, we
changed oscillator strengths.  We also introduced a handful of
artificial Fe~{\sc i} lines to mimic observed lines that did not
appear in our original line list.  We avoided changing molecular line
strengths.  Table~\ref{tab:linelist} gives the final line list that we
used in our computation of synthetic spectra
(Section~\ref{sec:synthesis}).

\subsubsection{Spectral Synthesis}
\label{sec:synthesis}

\begin{deluxetable}{lccc}
\tablewidth{0pt}
\tablecolumns{4}
\tablecaption{Grid of Synthetic Spectra\label{tab:grid}}
\tablehead{\colhead{Parameter} & \colhead{Start} & \colhead{End} & \colhead{Step}}
\startdata
$\lambda$           & 4100~\AA & 4500~\AA & 0.02~\AA \\
$T_{\rm eff}$       & 3500~K   & 5600~K   & 100~K \\
                    & 5600~K   & 6400~K   & 200~K \\
$\log g$            & 0.0      & 4.0      & 0.5   \\
$\feh$      & $-4.0$   & $0.0$    & 0.1   \\
${\rm [\alpha/Fe]}$ & $-0.8$   & $+1.2$   & 0.1   \\
$\cfe$      & $-2.4$   & $+1.0$   & 0.2   \\
                    & $+1.0$   & $+3.0$   & 0.4   \\
                    & $+3.0$   & $+3.5$   & 0.5   \\
\enddata
\end{deluxetable}

We used MOOG with our existing grid of ATLAS9 model atmospheres
\citep{kir11} to compute synthetic spectra at a range of atmospheric
parameters.  Table~\ref{tab:grid} shows the parameters of the spectral
grid: the lower and upper ranges and step sizes of wavelength,
effective temperature, surface gravity, metallicity, alpha
enhancement, and carbon abundance.  We used 256 CPUs to compute the
4,835,376 spectra.  Although [C/Fe] varied in the spectral syntheses,
the input model atmospheres were all computed assuming $\cfe =
0$.

As stars ascend the RGB, they dredge up products of the CNO cycle.  We
examine this phenomenon in Section~\ref{sec:astration}, but many
others have investigated it before.  In particular, \citet{kel01}
showed how the $^{12}$C/$^{13}$C isotope ratio varies as a function of
evolutionary state.  The following relation, which we deduced from
\citeauthor{kel01}'s Figure~4, approximates how the ratio changes with
decreasing surface gravity:

\begin{eqnarray}
\begin{array}{ll}
^{12}{\rm C}/^{13}{\rm C} = 50 & ~~~{\rm if}~\log g > 2.7 \\
^{12}{\rm C}/^{13}{\rm C} = 63\,\log g - 120 & ~~~{\rm if}~2.0 < \log g \le 2.7 \\
^{12}{\rm C}/^{13}{\rm C} = 6 & ~~~{\rm if}~\log g \le 2.0 \\ \label{eq:cratio}
\end{array}
\end{eqnarray}

\noindent
We used Equation~\ref{eq:cratio} in computing the grid of synthetic
spectra.

\subsection{Carbon Abundance Measurements}
\label{sec:abundance}

We measured carbon abundances by comparing the observed spectra to our
new grid.  This process involved several steps.

\begin{figure}[t!]
\centering
\includegraphics[width=\columnwidth]{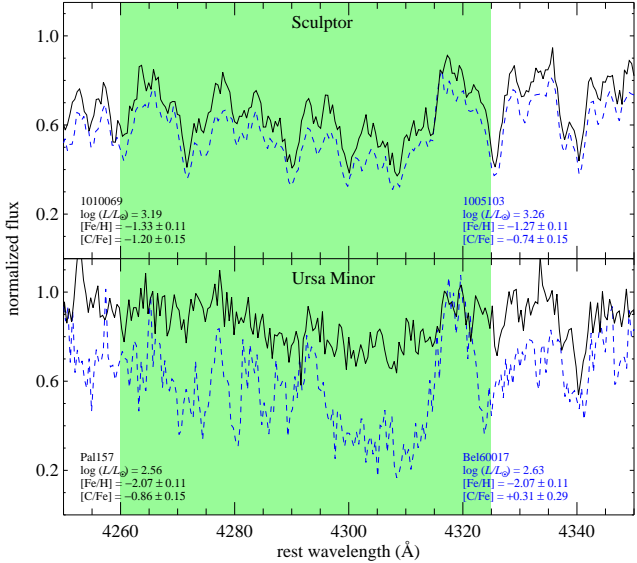}
\caption{G band spectra of pairs of stars with similar luminosity,
  similar metallicity, and differing carbon abundance in Sculptor
  (top) and Ursa Minor (bottom).  The stellar parameters listed on the
  left (right) of each panel correspond to the solid black (dashed
  blue) spectrum.  As in Figure~\ref{fig:gband}, only the green shaded
  region is used in the measurement of
  [C/Fe].\label{fig:gbandcompare}}
\end{figure}

\begin{enumerate}

\item {\it Continuum division:} MOOG generates synthetic spectra
  normalized to the continuum.  We divided each observed spectrum by a
  continuum fit so that it could be compared to the synthetic spectra.
  We fit the spectrum with a spline with a breakpoint spacing of
  200~\AA\@.  Pixels with fluxes that were greater than 5 standard
  deviations above or 0.1 standard deviations below the spline were
  iteratively excluded from the fit so that stellar absorption was not
  included in the continuum determination.  The spectrum was divided
  by the spline.  This continuum was refined in step~3.

\item {\it Preliminary [C/Fe] measurement:} We searched the spectral
  grid for the best-fitting value of [C/Fe].  The synthetic spectra
  were smoothed through a Gaussian kernel (2.1~\AA\ FWHM) to match the
  observed resolution.  We fixed $T_{\rm eff}$, $\log g$, [Fe/H], and
  [$\alpha$/Fe] at the values previously determined from the red
  DEIMOS spectra (see Section~\ref{sec:atm}).  We used the
  Levenberg--Marquardt optimization code MPFIT \citep{mar12} to find
  the value of [C/Fe] that minimized $\chi^2$ in the spectral range
  4260--4325~\AA\ (green region in Figures~\ref{fig:gband} and
  \ref{fig:gbandcompare}), where the spectrum is most sensitive to
  changes in carbon abundance.

\item {\it Continuum refinement:} The continuum determination in
  step~1 does not take into account the known variation in flux as a
  function of wavelength.  We refined the continuum by dividing the
  observed spectrum by the best-fitting synthetic spectrum determined
  in step~2.  We fit that quotient with a spline with a breakpoint
  spacing of 150~\AA\@.  This spacing is slightly finer than in step~1
  for a slightly more detailed continuum determination.  We
  iteratively discarded pixels that were above or below the fit by
  more than one standard deviation.

\item {\it Refined [C/Fe] measurement:} We repeated steps~2 and 3
  until [C/Fe] changed by less than 0.001 between iterations.

\end{enumerate}

In some cases, the spectra allowed only measurements of upper limits.
We computed $\chi^2$ contours for each star at several values of
[C/Fe] around the minimum $\chi^2$.  For stars whose $\chi^2$ contours
did not rise by at least 1 on both sides of the minimum, we calculated
a $2\sigma$ upper limit.  The upper limit was the value of [C/Fe] at
which $\chi^2$ was 4 above the minimum $\chi^2$.

Figure~\ref{fig:gband} shows model spectra of the best-fitting G band
spectrum in red.  Also shown in light blue are model spectra deviant
from the best fit at $\cfe \pm 0.2$.  Figure~\ref{fig:gbandcompare}
shows two pairs of spectra of stars---one pair in Sculptor and one
pair in Ursa Minor---with similar luminosity and metallicity but
different carbon abundances.  The figure demonstrates that the stars
measured to have larger [C/Fe] have noticeably stronger G bands.

We also computed the $S_2({\rm CH})$ index \citep{mar08a}.  This index
is defined to have minimal sensitivity to nitrogen abundance.  It is
the ratio of flux in the G band to flux in two side bands.  The blue
side band starts at 4212~\AA\@.  We measured $S_2({\rm CH})$ only in
spectra with minimum rest wavelengths less than 4212~\AA\@.
\citet{mar08a} defined the band to be measured in units of ADU per
pixel.  However, the throughput of DEIMOS in the vicinity of the G
band diminishes rapidly toward the blue.  Consequently, we measured
$S_2({\rm CH})$ in continuum-corrected flux rather than the original
ADU per pixel.

\addtocounter{table}{1}

Table~\ref{tab:cfe} gives the [C/Fe] and $S_2({\rm CH})$ measurements
for our GC and dSph sample.  The last column identifies the
\ncstar~carbon stars for which we were unable to measure [Fe/H] or
       [C/Fe].  The table also gives photometric information,
       including the absolute luminosity for each star relative to the
       solar luminosity.  We used bolometric corrections based on
       theoretical isochrones \citep{dem04,gir04} and previously
       measured distance moduli to compute the luminosities: 19.83 for
       NGC~2419 \citep{har97}, 15.21 for M68 \citep{mcc87}, 15.08 for
       M15 \citep{dur93}, 19.67 for Sculptor \citep{pie08}, 20.72 for
       Fornax \citep{riz07}, 19.18 for Ursa Minor \citep{mig99}, and
       19.84 for Draco \citep{bel02}.  The table also includes $T_{\rm
         eff}$, $\log g$, [$\alpha$/Fe], and S/N, calculated as the
       median absolute deviation of the observed spectrum from the
       best-fitting spectrum in the range 4250--4350~\AA\@.

\subsubsection{Estimation of Uncertainties}
\label{sec:uncertainty}

Although MPFIT provides a covariance matrix for each minimization,
this estimate of uncertainty on [C/Fe] does not account for
uncertainty in the placement of the continuum.  Instead of using the
covariance matrix, we calculated Monte Carlo uncertainties on [C/Fe]
by resampling the observed spectrum.

After we determined the best-fitting carbon abundance, we added noise
to the spectrum after it was continuum-normalized according to step~1
in Section~\ref{sec:abundance}.  The new flux in each noise-added
pixel was the original flux plus a value drawn from a Gaussian random
distribution with a standard deviation equal to spec2d's estimate of
the flux error in that pixel.  We repeated steps 2--4 above with the
noise-added spectrum.

We repeated this process for a total of 1000 noise realizations.  We
took the base uncertainty on [C/Fe] as the standard deviation of all
1000 trials.  Although this procedure accounts for random uncertainty,
it does not include all sources of systematic error.  We added 0.1~dex
as an estimate of systematic error in quadrature with the Monte Carlo
uncertainty for the final uncertainty on [C/Fe].  This systematic
uncertainty is approximately the value that is required to account for
the rms of the distribution of our measurements of [C/Fe] versus
literature measurements (Figure~\ref{fig:hires}, discussed in
Section~\ref{sec:validation}).

We computed errors on $S_2({\rm CH})$ from the same noise-added
spectra.  We did not add any systematic error to these measurements.

\subsubsection{Effect of Nitrogen and Oxygen}
\label{sec:no}

The measurement of carbon could be affected by the abundances of other
elements.  First, other elements that form diatoms with carbon affect
the molecular equilibrium, changing the amount of carbon available to
form CH\@.  Second, some elements contribute to the opacity in the
stellar atmosphere.  Changes in their abundances can affect the
structure of the atmosphere and the strength of the G band.  The most
important elements to consider for these effects are nitrogen and
oxygen because they are abundant, and they form diatoms with carbon.
Unfortunately, we can measure neither element directly from our
spectra.  However, we can quantify their possible effects on our
carbon abundance measurements.

We explored the effects of molecular equilibrium between CH and CN by
making syntheses with MOOG with varying abundances of N\@.  We took
three stars in M15 ($\feh \approx -2.4$) as test cases: 13701, which
is below the RGB luminosity function bump; 43026, which is at the top
of the RGB; and \gccrichname, which is carbon-enhanced (see
Section~\ref{sec:astration}).  We also tested two stars in Fornax to
explore more metal-rich ($\feh \approx -1$) stars: 52752 at $\log
(L/L_{\sun}) = 2.0$ and 40828 at the top of the RGB\@.  We synthesized
spectra with [N/Fe] values of $-2.0$, $-1.0$, $-0.5$, $-0.2$, $0.0$,
$+0.2$, $+0.5$, $+1.0$, and $+2.0$.  In no case did the flux of any
pixel in the G band change by more than 1\%.  

\begin{figure}[t!]
\centering
\includegraphics[width=\columnwidth]{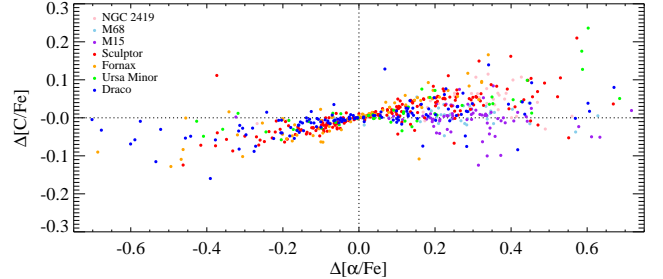}
\caption{The response of [C/Fe] to changing [$\alpha$/Fe] in the
  stellar atmosphere.  The $x$-axis shows the difference between the
  [$\alpha$/Fe] values used in our fiducial measurements of [C/Fe] and
  $[\alpha/{\rm Fe}] = 0.0$.\label{fig:alphafetest}}
\end{figure}

Our carbon abundance measurements already account for varying
elemental abundances that might affect the structure of the stellar
atmospheres.  We previously measured four $\alpha$ elements (Mg, Si,
Ca, and Ti) from red DEIMOS spectra \citep{kir10}.  The stellar
atmospheres we used to measure carbon abundances reflect those
[$\alpha$/Fe] ratios.  \citet{van12} showed that at temperatures found
in the atmospheres of cool giants, Mg and Si are the most important
metals for determining the opacity.  As a result, changing
[$\alpha$/Fe] (including [Mg/Fe] and [Si/Fe]) can alter the
atmospheric structure and therefore affect the measurements of all
elemental abundances.

We recomputed all of the [C/Fe] measurements assuming $[\alpha/{\rm
    Fe}] = 0.0$ rather than the values previously determined by
\citet{kir10}.  Here, $\alpha$ refers to O, Ne, Mg, Si, S, Ar, Ca, and
Ti.  This value of [$\alpha$/Fe] is used to compute both the model
atmosphere and the spectral synthesis.  Hence, the syntheses include
the effects of the varying atmospheric structure and the changing
molecular equilibrium between CH and CO\@.
Figure~\ref{fig:alphafetest} shows the response of [C/Fe] to the
change in [$\alpha$/Fe].  Changing [$\alpha$/Fe] by $\pm
0.6$~dex---much more than the mean uncertainty in [$\alpha$/Fe],
0.2~dex---results in a change of approximately $\pm 0.1$~dex in
[C/Fe].  Forcing [$\alpha$/Fe] to be 0.0 results in $|\Delta {\rm
  [C/Fe]}| < 0.1$~dex for 94\% of the dSph stars and $|\Delta {\rm
  [C/Fe]}| < 0.05$~dex for 75\% of the dSph stars.  We conclude that
uncertainty in [$\alpha$/Fe] is not a major source of uncertainty in
our carbon abundance measurements.

In conclusion, uncertainties in nitrogen, oxygen, and other $\alpha$
element abundances are minor contributors to the error budget of our
[C/Fe] measurements.

\subsubsection{Validation}
\label{sec:validation}

\begin{figure*}[t!]
\centering \includegraphics[width=\textwidth]{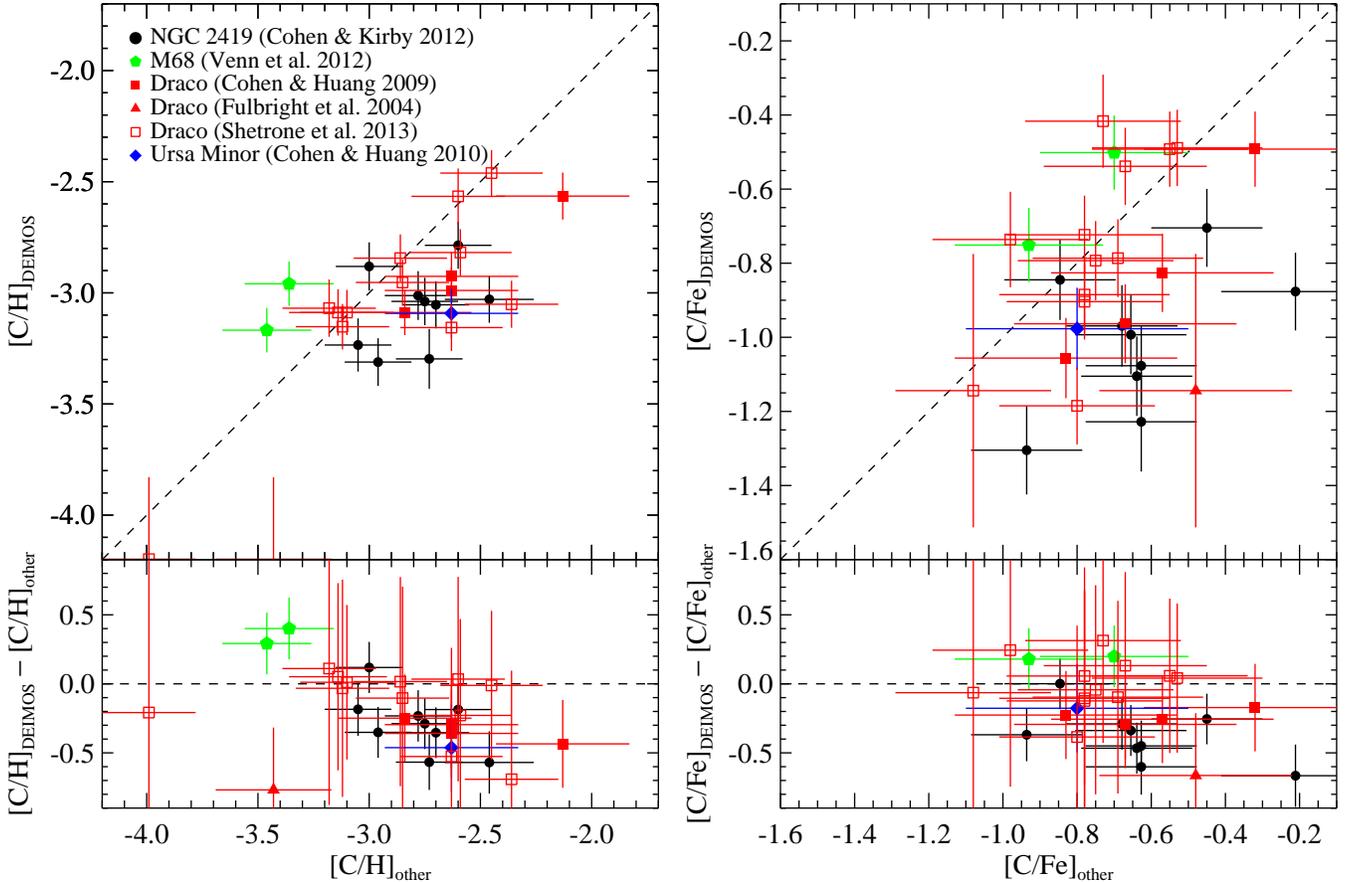}
\caption{Comparison between Keck/DEIMOS and literature carbon
  abundances for stars in the GCs NGC~2419 and M68 and the dSphs Draco
  and Ursa Minor.  All of the literature measurements come from
  Keck/HIRES except for the studies of \citet[][Magellan/MIKE]{ven12}
  and \citet[][Keck/LRIS]{she01}.\label{fig:hires}}
\end{figure*}

We compared our low-resolution measurements of [C/Fe] with previous
high- and low-resolution measurements of [C/Fe] for the same stars.
We found overlapping stars with high-resolution measurements in the
literature for NGC~2419 \citep{coh12}, M68 \citep{ven12}, Draco
\citep{ful04,coh09}, and Ursa Minor \citep{coh10}.  \citet{she13} also
published measurements of [C/Fe] in Draco based on low-resolution
Keck/LRIS spectroscopy.  All of these measurements are also based on
the G band.

The left panel of Figure~\ref{fig:hires} shows the comparison between
[C/H] measured from DEIMOS and [C/H] as published in the
aforementioned references.  The right panel shows [C/Fe].  We shifted
all abundances to our adopted solar abundance scale\footnote{$A({\rm
    X}) = 12 + \log (n({\rm X})/n({\rm H}))$ where $n({\rm X})$ is the
  number density of element X\@.}: $A({\rm C}) = 8.56$ \citep{and89}
and $A({\rm Fe}) = 7.52$ \citep{sne92}.  The deviations reach up to
0.8~dex.  The deviations are also about the same for [C/Fe] as they
are for [C/H], which indicates that differences in atmospheric
parameters like $T_{\rm eff}$ are not the reason for the discrepancy.
On the other hand, the degree of deviation depends largely on the
source of the measurement.  For example, our [C/H] measurements are
0.3--0.4~dex above \citeauthor{ven12}'s [C/H] measurements from
VLT/FLAMES, but our measurements are on average $\sim 0.4$~dex below
\citeauthor{coh09}'s, \citeauthor{coh12}'s, and \citeauthor{ful04}'s
measurements from Keck/HIRES\@.  Our measurements largely agree with
\citeauthor{she13}'s Keck/LRIS measurements.

The origin of the discrepancies might be spectral resolution or the
details of the carbon abundance measurements.  At first glance, the
spectral resolution is suspect because our low-resolution measurements
agree most with another low-resolution study \citep{she13}.  However,
our measurements disagree with \citeauthor{ven12}'s (\citeyear{ven12})
in the opposite direction from the other two high-resolution studies.
Therefore, we investigated the possibility that differences in the
line list caused the offset.  We compared our line list
(Table~\ref{tab:linelist}) with J.~Cohen's line list.  Between
4274~\AA\ and 4232~\AA, our list contains 2.5~times more CH lines.
The higher density of lines makes the G band stronger.  Hence, for a
fixed observed G band strength, we would measure a lower carbon
abundance with our line list than with Cohen's line list.  Our line
list has a similar density of lines to \citeauthor{she13}'s list.
Therefore, it is plausible that a different line list is the main
source of disagreement of carbon abundance measurements from the G
band.


\section{Astration of Carbon on the Upper RGB}
\label{sec:astration}

\begin{figure}[t!]
\centering
\includegraphics[width=\columnwidth]{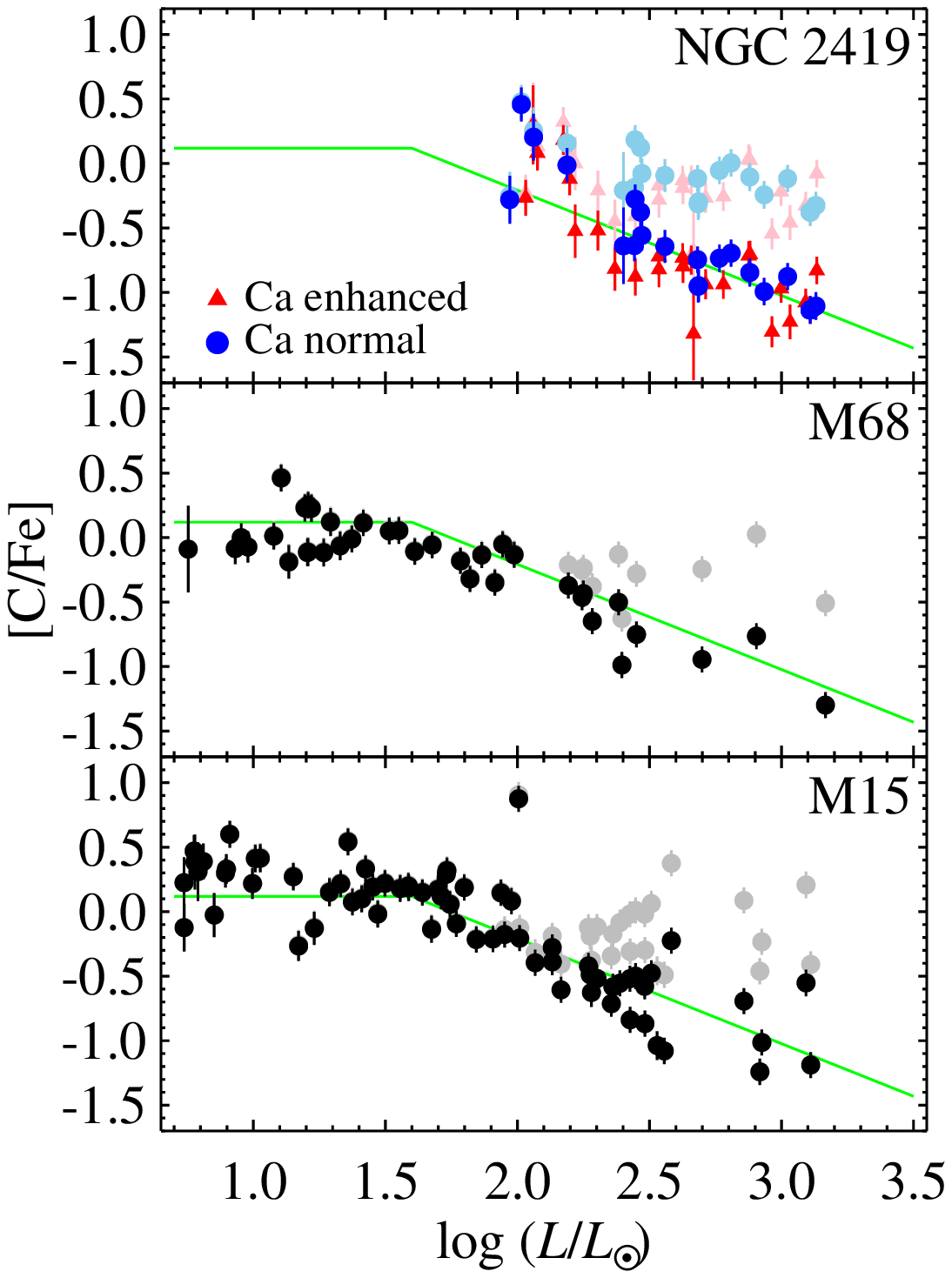}
\caption{The decrease in carbon enhancement with red giant luminosity
  in two globular clusters.  The giants in NGC~2419 are separated into
  calcium-enhanced (red triangles) and normal-calcium groups (red
  circles).  The green lines illustrate of the destruction of carbon
  as photospheric material is processed by the CNO cycle on the upper
  RGB (Equation~\ref{eq:cfetrend}).  The faded points show the [C/Fe]
  measurements corrected for the destruction of carbon with increasing
  luminosity according to the calculations of
  \citet{pla14}.\label{fig:gccfe}}
\end{figure}

\begin{figure*}[t!]
\centering
\includegraphics[width=1.7857\columnwidth]{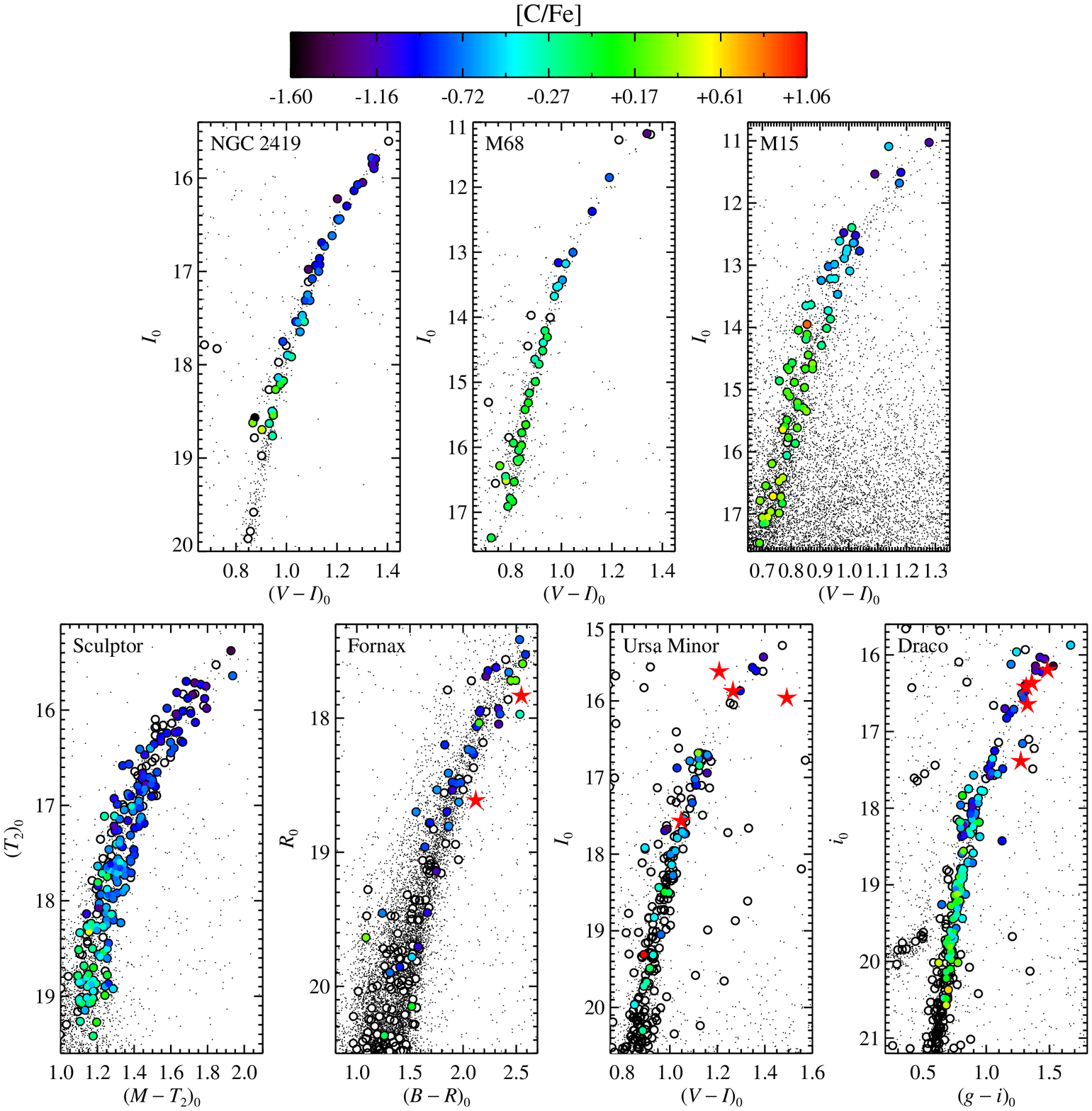}
\caption{Color--magnitude diagrams for the GCs and dSphs.  Point color
  corresponds to [C/Fe].  Hollow circles are spectroscopic targets
  that are non-members or spectra where \cfe\ was unable to be
  measured.  Red, five-pointed symbols indicate very carbon rich stars
  for which we did not attempt to measure \cfe.\label{fig:cmd}}
\end{figure*}

The destruction of carbon on the RGB above the luminosity function
bump is well documented \citep[e.g.,][]{sun81,car82,smi06}.  As
previous investigators have done, we searched for signatures of carbon
astration in GCs.  GCs are the easiest stellar systems to isolate
stellar luminosity as an independent variable because the cluster
stars have similar ages, metallicities, and distances.

Figure~\ref{fig:gccfe} shows the trend of carbon enhancement ([C/Fe])
with luminosity.  The carbon abundance remains flat until giants reach
$\log (L/L_{\sun}) = 1.6$, whereupon [C/Fe] declines, presumably due
to mixing that begins at the luminosity function bump.  We fit a line
to the combined trend of [C/Fe] with [Fe/H] for all three GCs, fixing
the luminosity of the bump at $\log (L/L_{\sun}) = 1.6$.  The best fit
line, represented in green in Figure~\ref{fig:gccfe}, is

\begin{eqnarray}
\begin{array}{ll}
\cfe = \cfebump \pm \cfeinterr & ~~~{\rm if}~\log (L/L_{\sun}) \le 1.6 \\
\cfe = (\cfeint \pm \cfeinterr) & \\
\hspace{0.5cm} - (\cfeslope \pm \cfeslopeerr)\,\log(L/L_{\sun}) & ~~~{\rm if}~\log (L/L_{\sun}) > 1.6 \\ \label{eq:cfetrend}
\end{array}
\end{eqnarray}

The amount of carbon astration in GCs could be computed from this
empirical trend, but we also considered theoretical estimates of the
amount of carbon depletion.  \citet{pla14} calculated corrections
based on models of mixing on the upper RGB\@.  The corrections depend
on surface gravity, metallicity, and uncorrected carbon
abundance.\footnote{The correction depends slightly on [N/Fe].  We
  assumed ${\rm [N/Fe]} = 0$.  \citet{gra00} showed that [N/Fe] climbs
  to about 0.5 at the tip of the RGB in metal-poor field stars.
  \citeauthor{pla14}'s (\citeyear{pla14}) corrections change by
  $<0.05$~dex for changes in [N/Fe] of 0.5~dex.  For the range of
  stellar parameters in our sample, the difference from the ${\rm
    [N/Fe]} = 0$ correction exceeds 0.1~dex only for ${\rm [N/Fe]} \ge
  +1$, a condition that very few, if any, of the stars in our sample
  would satisfy.}  We used their models to compute a correction for
each star in our sample.  Figure~\ref{fig:gccfe} shows in faded colors
\citeauthor{pla14}'s corrections applied to the GC sample.  The faded
points have less dependence on luminosity than the uncorrected points.
Although the empirical linear fit to the data
(Equation~\ref{eq:cfetrend}) is a better fit than \citeauthor{pla14}'s
models, the models are ab initio predictions.  They can be applied to
data in environments that are not as tightly controlled as GCs.  We
use the \citet{pla14} corrections for dSph stars in
Section~\ref{sec:dsphs}.

\citet{mar08c} found that the rate of carbon depletion as giants
evolve depends on metallicity, which is consistent with the
thermohaline mixing mechanism for the change in elemental composition
with luminosity \citep{cha07,cha10}.  Unfortunately, we can not test
for metallicity dependence because the GCs we observed have similar
metallicities.  According to \citet{har96}, the metallicities are
$\feh = -2.15$ \citep[NGC~2419,][]{sun88}, $-2.23$
\citep[M68,][]{lee05}, and $-2.37$ \citep[M15,][]{tak09}.
\citeauthor{mar08c}\ found carbon depletion rates of $d\cfe/d\log
(L/L_{\sun}) = -0.7$ to $-1.5$ above the luminosity function bump for
the metallicity range of these three GCs.  Our value of $-\cfeslope
\pm \cfeslopeerr$ (Equation~\ref{eq:cfetrend}) falls in this range.

GCs have primordial abundance variations \citep{gra04}, including
carbon \citep[e.g.,][]{coh99b}.  The dispersion in [C/Fe] at fixed
luminosity can give a sense of the differential enrichment in carbon
between the first and subsequent generations of star formation.  We
calculated this dispersion by finding the residuals between [C/Fe] and
the kinked linear trend.  The trend is similar to
Equation~\ref{eq:cfetrend}, but we fit each GC individually to
calculate the dispersion.  We then normalized the residuals by the
measurement uncertainty: $\Delta = (\cfe - \cfe({\rm trend})) /
\delta\cfe$.  If there is no intrinsic scatter in [C/Fe] beyond the
measurement uncertainty, then the standard deviation of $\Delta$
should be 1.  Instead, we found \deltaa\ for NGC~2419, \deltab\ for
M68, and \deltac\ for M15.  (We excluded star \gccrichname, discussed
below, from M15.)  This measurement of dispersion demands that we
estimated measurement uncertainties accurately.  If we underestimated
the uncertainties, then the standard deviation of $\Delta$ will appear
erroneously larger than 1.  Taken at face value, we have found
evidence for intrinsic scatter in [C/Fe].  We see no evidence for
bimodality in [C/Fe], but the scatter is not much larger than the
uncertainties.  Nonetheless, CN bands do show bimodality in some GCs
\citep{smi05,smi06,mar08b}, reflecting that primordial C and N
abundances in clusters were bimodal, independent of the present-day
astration on the upper RGB\@.  This bimodality can even be found on
the main sequence of some clusters \citep[e.g., 47 Tucanae,][]{har03}.

\citet{coh11}, \citet{coh12}, and \citet{muc12} discovered that
NGC~2419 has an extremely unusual chemical composition.  The Mg
abundances vary from star to star by a factor of 25, and the variation
is anticorrelated with abundances of other elements, including K and
Ca.  In a sample of 13 red giants with high-resolution spectra,
\citet{coh12} did not find any correlation between C and any of the
elements with an intracluster dispersion.  We looked in our larger
sample for such a correlation by separating the NGC~2419 stars into
Ca-enhanced and Ca-normal.  We estimated Ca enhancement by measuring
the equivalent widths (EWs) of the near-infrared Ca triplet.  We
converted these EWs to metallicity estimates using
\citeauthor{sta10}'s (\citeyear{sta10}) formula.  The stars with
$\feh_{\rm CaT} < -2.2$ (Ca-normal) are represented by blue circles in
Figure~\ref{fig:gccfe}.  Stars with $\feh_{\rm CaT} \ge -2.2$
(Ca-enhanced) are red triangles.  There is no noticeable difference
between the two populations.  Hence, we confirm that carbon did not
participate in the unusual nucleosynthesis that affected heavier
elements in NGC~2419.

Figure~\ref{fig:cmd} shows an alternate way to display the destruction
of carbon as red giants ascend the RGB\@.  The CMD color-codes the
stars, with cooler colors corresponding to lower [C/Fe].  In general,
the plotting color is cooler at the top of the RGB compared to less
luminous stars.  In M68 and M15, which are the GCs observed to
sufficient depth, most of the stars below the luminosity function bump
($I_0 \sim 14$) are colored green.  This color corresponds to the
constant pre-bump value of $\cfe = +\cfebump$.

Note the one carbon-rich star in M15, which is also visible in
Figure~\ref{fig:gccfe}.  Star \gccrichname, a confirmed radial
velocity member of M15, stands out as lone orange point ($\cfe =
\gccrichcfe \pm \gccrichcfeerr$) in Figure~\ref{fig:cmd} in a sea of
dark green points ($\cfe \sim -0.2$).  The carbon abundance is high
enough that the near-infrared spectrum shows some CN absorption
(Figure~\ref{fig:cstars}).  This star was also previously identified
by J.~Cohen to be carbon-rich from an unpublished Keck/HIRES spectrum.
That spectrum shows a weak C$_2$ bandhead at 5163~\AA\ in addition to
a very strong G band.  Also note that this star lies slightly blueward
of the main RGB locus.  The CMD position indicates that star
\gccrichname\ is an AGB star.

Carbon-enhanced stars in GCs are rare.  \citet{coh78} and
\citet{mou02} explain that C stars appear after several hundred
million years, long after the end of star formation in a GC\@.  None
of the GC stars would have been pre-enriched by carbon from an AGB
star.  Additionally, the lowest-mass AGB star that produces copious
amounts of carbon is about $2~M_{\sun}$.  Such a star would live for
about 1~Gyr.  Hence, all carbon-producing AGB stars in GCs should have
died long ago.  Nonetheless, there are a few examples of C-rich stars
in GCs.  The GC $\omega$ Centauri hosts at least seven known C stars
\citep{har62,dic72,bon75,cow85,van07}.  However, $\omega$ Centauri is
known to have an unusually long star formation history (SFH) for a GC
\citep{hil04,vil07}.  Regardless, C stars can be found in even more
ordinary GCs, like M14 \citep{cot97}, Lynga~7 \citep{mat06,fea13}, and
NGC~6426 \citep{sha12}.

C stars can be formed as the result of mass transfer from an AGB star.
In most cases, the AGB star has now evolved into a white dwarf.  There
is photometric \citep{boh00} and spectroscopic \citep{luc05,sta14}
evidence that most or all C stars---especially those enhanced in
$s$-process elements---have binary companions.  Binarity is often
diagnosed through radial velocity variability over multiple
measurements.  Although we observed \gccrichname\ twice, the
observations were separated by only 24~hours.  We found no evidence
for binarity from this very short time baseline.  An alternative
possibility is that the existing C-rich AGB star formed in a stellar
merger, the result of which would be massive enough to generate carbon
in its AGB phase \citep[e.g.,][]{fea13}.  Stellar mergers are a
plausible origin for blue stragglers in GCs \citep{mat90,sil13}, so it
is conceivable that this carbon-rich star is also the product of a
merger.


\section{Carbon in Dwarf Spheroidal Galaxies}
\label{sec:dsphs}

Unlike GCs, dSphs contain stars with a range of ages and
metallicities.  The more complex stellar populations complicate the
evolution of carbon.  First, dSph stars need not start their lives
with the same C abundance.  GC stars were formed from well-mixed gas
polluted only by Type~II supernovae (SNe) and AGB stars in a limited
mass range.  Figure~\ref{fig:gccfe} and Equation~\ref{eq:cfetrend}
show that all of the pre-bump stars within a GC have roughly the same
value of $\cfe \sim +\cfebump$.  On the other hand, the progenitors of
dSph stars can have a range of metallicities.  They can also form from
inhomogeneously mixed gas polluted by SNe of various types and AGB
stars of various masses and metallicities.  Second, the star formation
duration of dSphs is long enough for 2--5~$M_{\sun}$ AGB stars---the
ones that generate copious amount of carbon---to pollute the
interstellar medium while the galaxy is still forming stars.  As a
result, AGB stars influence the evolution of carbon differently in
dSphs than in GCs.  Third, the stellar density of dSphs is
significantly lower than GCs, which reduces the stellar merger rate.
Consequently, the merger formation channel for prematurely C-rich
stars (possibly like \gccrichname\ in M15) is suppressed.

\begin{figure*}[p!]
\centering
\includegraphics[width=1.6\columnwidth]{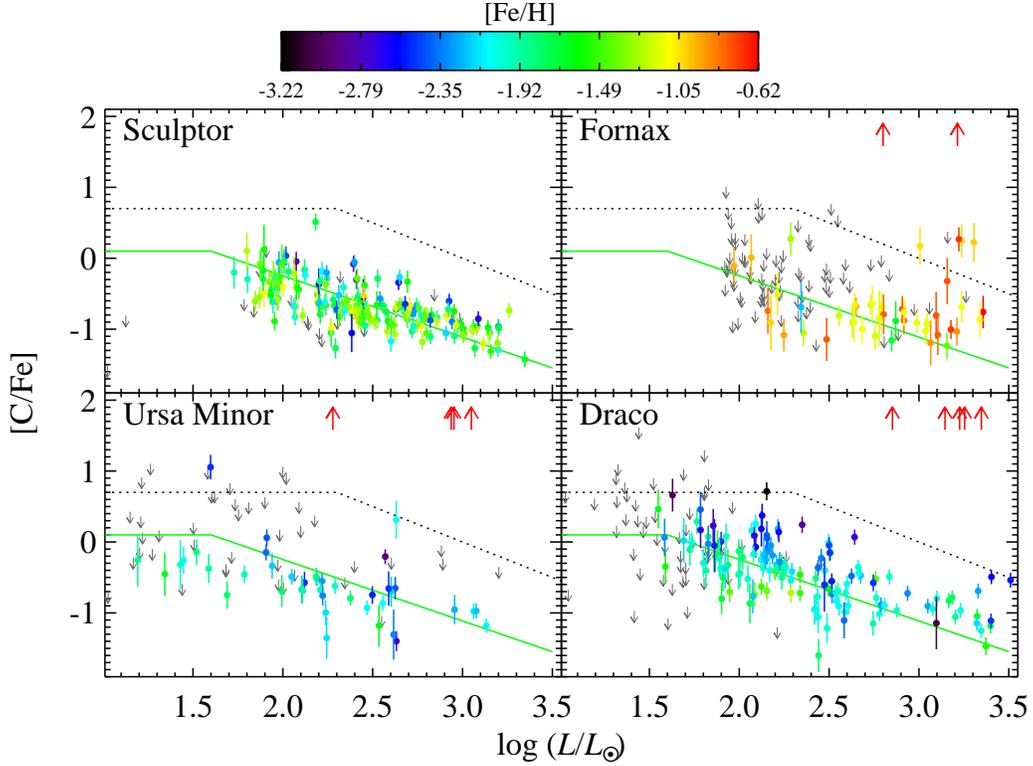}
\caption{Carbon enhancement as a function of stellar luminosity in
  dSph stars.  The green line (Equation~\ref{eq:cfetrend}), copied
  from the GC data in Figure~\ref{fig:gccfe}, shows the destruction of
  carbon as photospheric material is processed through the CNO cycle
  on the upper RGB\@.  The color coding corresponds to iron abundance.
  The dotted line is \citeauthor{aok07}'s (\citeyear{aok07}) dividing
  line for carbon enhancement.  Upward-pointing red arrows indicate
  very carbon rich stars for which we did not attempt to measure \cfe.
  Downward-pointing gray arrows are $2\sigma$ upper
  limits.\label{fig:dsphcfelogl}}
\end{figure*}

\begin{figure*}[p!]
\centering
\includegraphics[width=1.6\columnwidth]{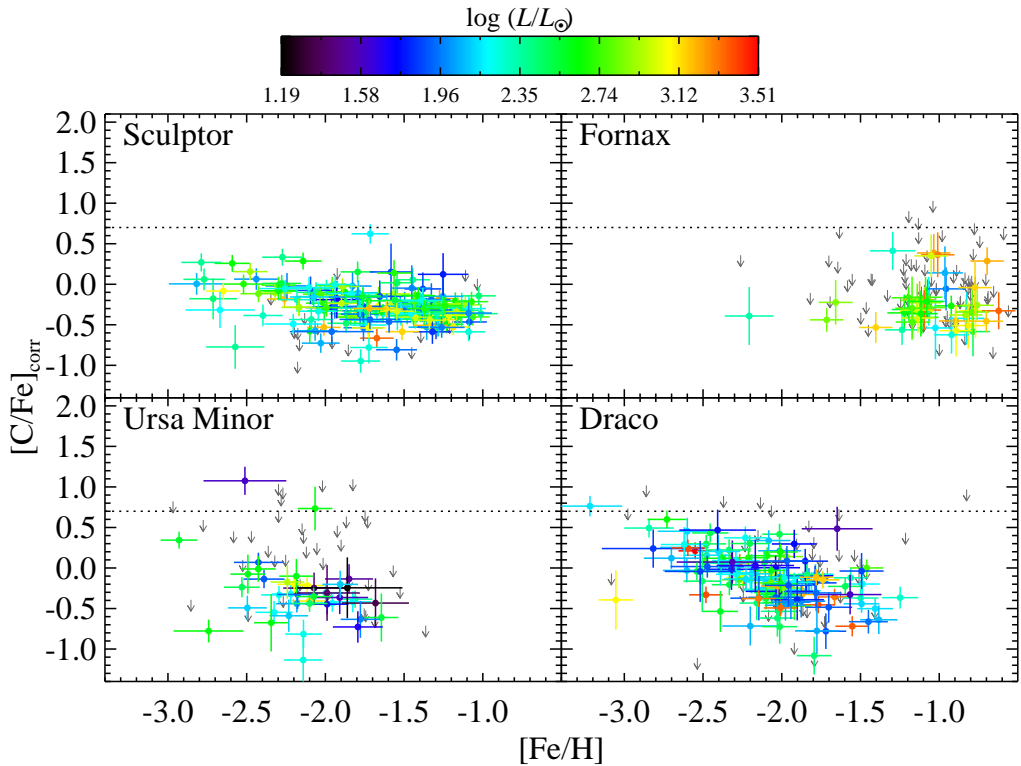}
\caption{Carbon enhancement corrected for luminosity according to
  \citet{pla14} as a function of stellar metallicity.  The color
  coding corresponds to stellar luminosity.  The dotted line is one of
  \citeauthor{pla14}'s definitions for carbon enhancement.
  Downward-pointing gray arrows are $2\sigma$ upper limits.  The C
  stars are not plotted because we did not measure [Fe/H] for those
  stars.\label{fig:dsphcfefeh}}
\end{figure*}

The bottom panels of Figure~\ref{fig:cmd} show that the decline of
[C/Fe] with increasing stellar luminosity is just as present in dSphs
as in GCs.  However, there is a larger spread in [C/Fe] at fixed
luminosity in the dSphs.  Figure~\ref{fig:dsphcfelogl} shows the
astration of carbon in dSphs, just as Figure~\ref{fig:gccfe} shows the
same for GCs.  The green lines in Figure~\ref{fig:dsphcfelogl} are the
same as in Figure~\ref{fig:gccfe}.  They were not re-fit to the dSph
data.  The slope of the line roughly tracks the decline of [C/Fe] in
the dSphs, especially in Sculptor and Draco.

The scatter in [C/Fe] at fixed luminosity exceeds the scatter expected
from measurement uncertainty.  The residuals are correlated with
[Fe/H], which is represented by the color coding in
Figure~\ref{fig:dsphcfelogl}.  The residual metallicity trend is
especially apparent in Ursa Minor and Draco, the most metal-poor of
the four dSphs we observed.  The stars with the lowest [Fe/H] have the
highest [C/Fe] at fixed luminosity.  This trend indicates that C
abundance did not increase as quickly as Fe abundance as star
formation progressed in the dSphs.  Later forming stars, which have
higher [Fe/H], started with lower [C/Fe] ratios than their
predecessors.

Figure~\ref{fig:dsphcfefeh} shows the trend in another way.  The
$x$-axis is [Fe/H] instead of luminosity.  Instead of [C/Fe], we plot
$\cfecorr$, which is corrected for the depletion due to astration on
the upper RGB according to \citet{pla14}.  This correction removes the
luminosity dependence so that we may isolate [Fe/H] as the independent
variable.  Figure~\ref{fig:dsphcfefeh} confirms that $\cfecorr$
declines with increasing [Fe/H] in Draco.  The trend is also apparent
in Ursa Minor and Sculptor, where stars with $\feh < -2.2$ have higher
$\cfecorr$ on average than stars with $\feh > -2.2$.  The sample of
stars in Fornax does not span a large enough range of [Fe/H] to draw
definitive conclusions on the metallicity dependence of $\cfecorr$.

We considered the possibility that $\cfecorr$ appears to decline with
[Fe/H] because our spectra lack the S/N or resolution to measure
carbon abundances at low metallicities.  We calculated upper limits
for all of the stars where we could not measure carbon abundances.  If
the $\cfecorr$ trend appeared because of the loss of measurability at
low [Fe/H], then the upper limits would trace the lower envelope of
$\cfecorr$ measurements.  Instead, the upper limits are distributed
fairly evenly in $\cfecorr$.  We conclude that the decline is real.

\begin{figure}[t!]
\centering
\includegraphics[width=\columnwidth]{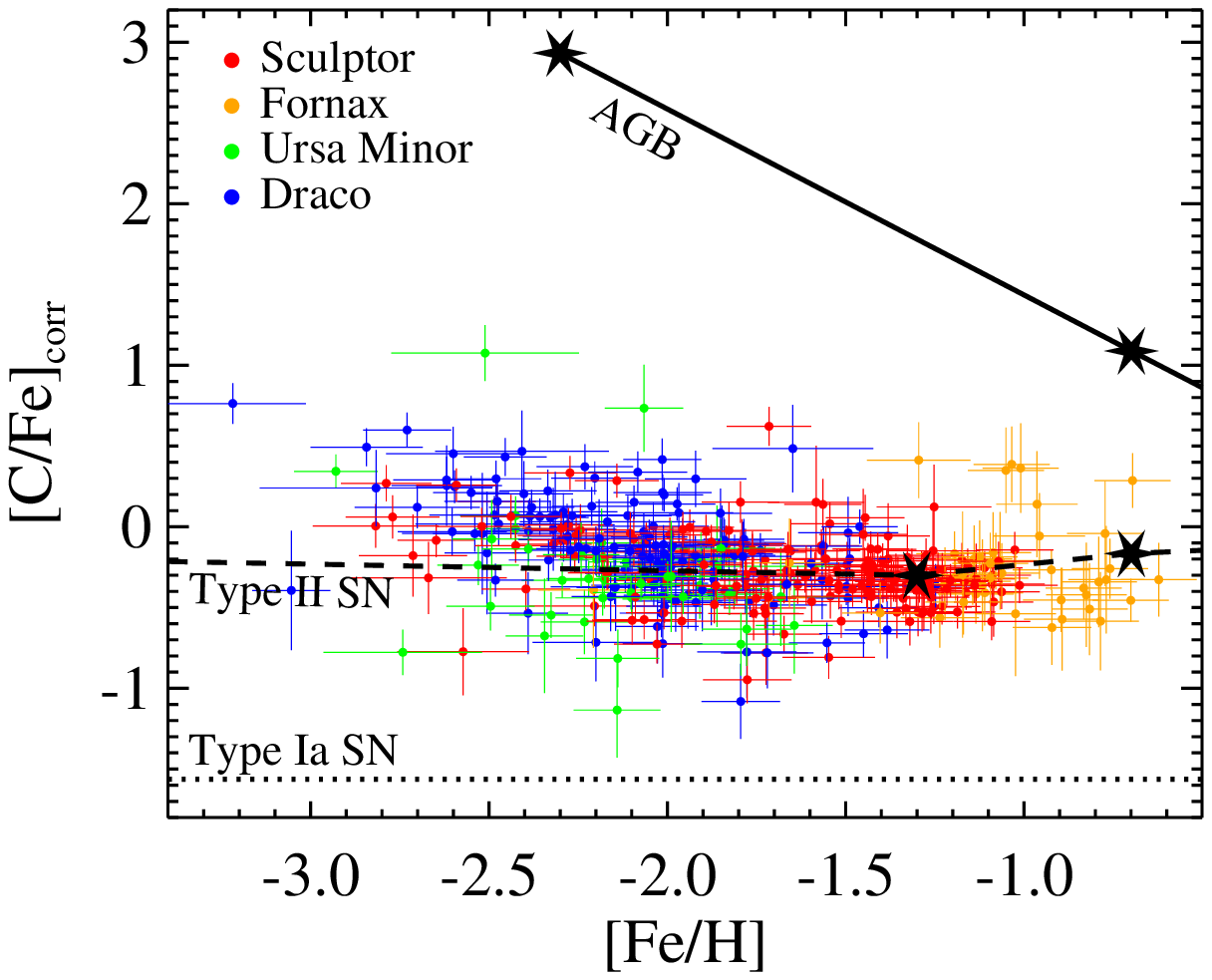}
\caption{Carbon enhancement corrected for luminosity according to
  \citet{pla14} for all four dSphs.  The theoretically predicted,
  IMF-averaged yields for nucleosynthetic sources are also shown: AGB
  stars \citep[solid line,][]{kar10}, Type~II SNe \citep[dashed
    line,][]{nom06}, and Type~Ia SNe \citep[dotted line,][]{iwa99}.
  The AGB and Type~II SN yields depend on metallicity.  AGB stars
  could be active at $\feh < -2.3$, but \citeauthor{kar10}'s models
  are not available at lower [Fe/H].  Six-pointed symbols indicate the
  metallicities at which the yields were computed.  A decline in
  $\cfecorr$ can be explained by the metallicity dependence of AGB
  yields combined with a growing contribution of Type~Ia
  SNe.\label{fig:nucleosynthesis}}
\end{figure}

The decline of $\cfecorr$ with [Fe/H] can be explained by the varying
contribution of different nucleosynthetic sources over time.
Figure~\ref{fig:nucleosynthesis} shows theoretical nucleosynthetic
yields superimposed on the same data as in
Figure~\ref{fig:dsphcfefeh}.  We adopted the AGB yields of
\citet{kar10}, the Type~II SN yields of \citet{nom06}, and the Type~Ia
SN yields of \citet[][model W7]{iwa99}.

Type~II SNe are the first sources to enrich a galaxy with C and Fe.
Figure~\ref{fig:nucleosynthesis} shows the Type~II SN yields averaged
over a \citet{sal55} initial mass function (IMF)\@.  In practice, the
most massive SNe explode first.  According to \citet{nom06}, a
$40~M_{\sun}$ SN has a C/Fe yield 5 times that of a $13~M_{\sun}$ SN
($\cfe = +0.36$ versus $-0.36$).  As a result, the very earliest stars
in the galaxy may have been able to incorporate the ejecta of only the
most massive, shortest lived SNe, which have the highest [C/Fe] of the
Type~II SNe.  The two stars in our sample with $\feh < -3$, both in
Draco, have $\cfecorr = \cfecorrempa \pm \cfeerrempa$ and
$\cfecorrempb \pm \cfeerrempb$.  The difference between these values
could indicate separate enrichment from Type~II SNe of different
masses, or it could indicate different levels of AGB pollution.

The IMF-averaged Type~II SN yields are lower than the observed [C/Fe]
in the majority of dSph stars at $\feh \la -1.7$.  These stars require
a more carbon-rich nucleosynthetic source, such as AGB stars.  AGB
stars were likely polluting the interstellar medium even at the very
lowest metallicities.  \citet{kar10} did not calculate AGB yields at
$\feh < -2.3$, but AGB stars surely existed at lower metallicities.
If the high [C/Fe] yields persist at $\feh < -2.3$, then AGB ejecta
could easily explain the high [C/Fe] values observed at low [Fe/H] in
the dSphs.  Their ejecta would mix with Type~II SN ejecta to produce
the [C/Fe] ratio observed in low-metallicity stars.  Because AGB stars
have much larger [C/Fe] ratios than Type~II SNe, it would not take
very many AGB stars compared to Type~II SNe to explain the carbon
measurements in low-metallicity dSph stars.

The AGB yields are highly metallicity-dependent.  As the galaxy's
metallicity increases, the AGB stars eject more and more Fe but about
the same amount of C\@.  From $\log (Z/Z_{\sun}) = -2.3$ to $-0.7$,
the C output decreases by a factor of 2 while the Fe output increases
by a factor of 37.  Hence, the metallicity dependence of AGB yields
could explain some of the decline in [C/Fe] with increasing [Fe/H].

Type~Ia SNe are the last nucleosynthetic sources to turn on in dSphs
because they are the end points of some long-lived, low-mass stars.
The Type~Ia [C/Fe] yield is $-1.6$, significantly below the yield of
Type~II SNe and AGB stars.  As the galaxy evolves, the ratio of
Type~Ia to Type~II SNe increases.  Because the abundance ratios of the
two SN types are drastically different, the trend of abundance ratios
with [Fe/H] reflects the changing contribution of the two SN types to
the galaxy's chemical evolution \citep[e.g.,][]{gil91}.  The elements
typically used to diagnose this ratio are $\alpha$ elements (O, Mg,
and Si among others) and Fe.  As Figure~\ref{fig:nucleosynthesis}
demonstrates, C/Fe is also sensitive to the Type~II-to-Type~Ia SN
ratio.  Although the heavy production of C in AGB stars complicates
the interpretation of the C/Fe ratio, some of the decline in [C/Fe]
with increasing [Fe/H] in dSphs must be due to the increasing
frequency of Type~Ia SNe with time.

The four dSphs show different [C/Fe] trends.  The [C/Fe] ratio
declines most steeply in Ursa Minor and Draco.  Under the presumption
that an increasing prevalence of Type~Ia SNe drives the decline in
[C/Fe], then these two dSphs would have had a very inefficient SFH\@.
The galaxy would have taken about 100~Myr---the approximate delay time
for a Type~Ia SN---to reach $\feh \approx -2.5$, the metallicity at
which [C/Fe] begins to decline.  Unfortunately, Ursa Minor, Draco, and
Sculptor are too old to permit fine time resolution in measurements of
their SFHs \citep[e.g.,][]{orb08}, so our chemical interpretation can
not be cross-checked against photometrically measured SFHs.  Finally,
Fornax has a complex SFH \citep[e.g.,][]{deb12}.  Its pattern of
[C/Fe] is unlike the other three dSphs.  Some of the most metal-rich
stars have high values of $\cfecorr$.  These measurements also have
some the largest uncertainties, largely because the spectrum at
4300~\AA\ is blanketed with atomic and molecular absorption lines.  If
these measurements are trustworthy, then we speculate that AGB stars
are responsible for the high carbon abundances, possibly through mass
transfer.

We conclude that the primary causes of the decline of $\cfecorr$ with
[Fe/H] are (1) the metallicity dependence of AGB yields and (2) the
increasing ratio of Type~Ia to Type~II SNe with time.  A quantitative
interpretation of Figures~\ref{fig:dsphcfefeh} and
\ref{fig:nucleosynthesis} would require a chemical evolution model.
\citet{kir11b} developed such a model to explain the decline of
      [$\alpha$/Fe] with [Fe/H] in dSphs, including the four dSphs
      discussed here.  One prospect for future work is to incorporate
      carbon into the model, paying close attention to the yields of
      AGB stars.

Metallicity does not completely explain the observed scatter in
[C/Fe].  \citet{gra00} observed a large scatter of [C/Fe] on the upper
RGB in low-metallicity field stars in the MW halo.  However, they did
not have a large enough sample of unevolved stars at low metallicity
to determine whether the scatter was imparted to the stars at birth or
if it was caused by star-to-star variations in the efficiency of
carbon astration on the upper RGB\@.  \citet{spi05} found an
approximately constant value of [C/Fe] in unevolved field stars
(ignoring C stars), even at low metallicities.  This suggests that
low-metallicity, star-forming gas in the early MW halo had a uniform
[C/Fe] ratio.  However, \citet{spi06} found that evolved stars on the
upper RGB did not all experience the same level of mixing.  In
particular, the [(C+N)/Fe] ratio was larger for evolved stars at $\feh
< -2.5$ than for higher-metallicity stars.  In CNO-processed material
that did not undergo much of the high-temperature ON cycle, the
[(C+N)/Fe] ratio should be constant.  Hence, \citet{spi06} uncovered a
metallicity dependence in the efficiency of mixing.

In principle, \citeauthor{pla14}'s (\citeyear{pla14})
metallicity-dependent carbon corrections should account for the
variation in mixing efficiencies found by \citet{spi06}.  However, we
still observe a dispersion in $\cfecorr$ beyond the measurement
uncertainty at fixed [Fe/H].  We fit a quadratic relation between
$\cfecorr$ and [Fe/H] to stars with $\cfecorr < +0.7$ (to exclude
stars that may have acquired carbon through binary mass transfer).  In
a manner similar to calculating the intrinsic dispersion of [C/Fe] in
GCs (Section~\ref{sec:astration}), we calculated the residuals,
$\Delta$, about the quadratic fit and divided them by the measurement
uncertainties: $\Delta = (\cfecorr - \cfecorr({\rm fit})) /
\delta\cfe$.  If measurement uncertainties explain the scatter, then
the standard deviation of $\Delta$ should be 1.  Instead, we found
standard deviations of \residscl\ for Sculptor, \residfor\ for Fornax,
\residumi\ for Ursa Minor, and \residdra\ for Draco.  Therefore, all
of the dSphs show a small residual beyond the measurement
uncertainties that can not be explained by the decline of [C/Fe] with
increasing stellar luminosity (Figure~\ref{fig:dsphcfelogl}) or with
increasing metallicity (Figure~\ref{fig:dsphcfefeh}).

One possible source of the residual scatter is inhomogeneous mixing
\citep[e.g.,][]{rev12}.  If different contemporaneous star formation
regions formed out of gas that was not mixed across regions, then
stars at the same metallicity and luminosity could have different
abundance ratios.  Also, the calculation of $\Delta$ depends on
accurate estimates of measurement uncertainty.  If we have
underestimated $\delta\cfe$, then we have also overestimated $\Delta$.
Another possibility is that \citeauthor{pla14}'s (\citeyear{pla14})
corrections are imperfect.  Errors on the order of $\sim 0.15$~dex in
the corrections could cause the scatter in $\cfecorr$ that we observe.

\subsection{Carbon Stars}
\label{sec:dsphcstars}

\begin{deluxetable*}{llll}
\tablewidth{0pt}
\tablecolumns{4}
\tablecaption{Carbon Stars\label{tab:cstar}}
\tablehead{\colhead{DSph} & \colhead{Name} & \colhead{Reference} & \colhead{Reference ID}}
\startdata
Fornax     & 50465    & this paper    & \nodata \\
Fornax     & 98788    & this paper    & \nodata \\
Ursa Minor & Bel60031 & \citet{aar83} & 34227   \\
           &          & \citet{arm95} & VA335   \\
Ursa Minor & Bel60230 & this paper    & \nodata \\
Ursa Minor & Bel80022 & \citet{she01} & UM~1545 \\
Ursa Minor & Bel10023 & \citet{can78} & COS215  \\
           &          & \citet{zin81} & K       \\
           &          & \citet{aar83} & K       \\
           &          & \citet{arm95} & K       \\
           &          & \citet{she01} & K       \\
           &          & \citet{win03} & COS215  \\
Draco      & 670092   & \citet{arm95} & 461     \\
Draco      & 569394   & \citet{aar82} & 3203    \\
           &          & \citet{arm95} & C1      \\
Draco      & 607050   & \citet{aar82} & J       \\
           &          & \citet{arm95} & C2      \\
Draco      & 571725   & \citet{aar82} & 3237    \\
           &          & \citet{arm95} & C3      \\
Draco      & 682522   & \citet{azz86} & 578     \\
           &          & \citet{arm95} & C4      \\
\enddata
\end{deluxetable*}

As demonstrated in M15 (Section~\ref{sec:astration}), stars need not
be born with their present [C/Fe] ratio.  They can acquire large
amounts of carbon through mass transfer from an AGB star.  These stars
can exhibit $\cfe \gg +1$.  We identified such stars in our dSph
sample.  We found \ncstarfor\ C stars in Fornax, \ncstarumi\ in Ursa
Minor, \ncstardra\ in Draco, and none in Sculptor.  They are
represented as red, five-pointed symbols in Figure~\ref{fig:cmd}.
Most of these stars were discovered in previous
surveys.\footnote{\citet{she01} identified star 622253 (called 68 in
  their paper) in Draco as a C star.  We do not find it to be
  carbon-rich.  We measured its carbon enhancement to be $\cfe =
  \shecstarcfe \pm \shecstarcfeerr$.}  Table~\ref{tab:cstar} lists all
of the C stars in our sample along with the the reference(s) and
identification name(s) for their previous discoveries, where
applicable.  They all have extremely strong CN bands, as shown in
Figure~\ref{fig:cstars}.  Most of the C stars also have very strong
C$_2$ bands visible in the blue spectra.  However, two of the three
Ursa Minor C stars have weak C$_2$, and C$_2$ is not visible at all in
star Bel80022.  It is likely too metal-poor and perhaps too warm for
C$_2$ to appear in our low-resolution spectrum.

The resolution of our DEIMOS spectra does not permit a detailed
chemical abundance analysis of these carbon stars.  The CN, C$_2$, and
CH molecular bands blanket the spectra too heavily.  For this reason,
their carbon abundances are represented as lower limits in
Figure~\ref{fig:dsphcfelogl}.  They are not represented at all in
Figure~\ref{fig:dsphcfefeh} because we could not measure [Fe/H].
Higher resolution spectra would allow a detailed chemical analysis
\citep[e.g.,][]{kar14}.  Almost all of the carbon stars we found are
in fact bright enough to be observed with a high-resolution
spectrograph on a large telescope, like Keck/HIRES\@.  We reserve this
task for future work.

Figure~\ref{fig:cmd} shows that five of the \ncstar\ C stars are
redward of the RGB\@.  The tendency for carbon stars (or more
specifically, barium stars, which almost always have excess carbon) to
appear red in optical colors is called the Bond--Neff effect
\citep{bon69}.  The source of the effect is a depression of blue flux
due to molecular blanketing of the spectrum.  In fact, all five of the
C stars redder than the RGB show strong C$_2$, CH, and CN\@.

Our spectroscopic target selection was focused on selecting stars from
the RGB\@.  Hence, we may have missed many C stars.  Although we
discuss the C stars in our sample, the sample is definitely not
complete.  Other surveys that specifically searched for C stars
\citep[e.g.,][]{dem02} are better able to quantify their frequency.
Our sample selection for Sculptor was especially biased against
finding C stars.  \citet{sku15} found the first CEMP star in Sculptor,
but it lies outside of the area we surveyed.

The brightest C star in Ursa Minor, Bel10023, is actually blueward of
the RGB\@.  The molecular absorption apparently is not strong enough
to redden the optical colors.  Indeed, the C$_2$ bands in this star
are visible but very weak.  The star is probably an AGB star, just
like \gccrichname\ in M15.  It possibly manufactured its carbon by
itself rather than acquiring it through binary mass transfer, though
the star is probably too low-mass to dredge up carbon.  Unfortunately,
our spectra do not provide any clues as to whether the C stars
generated their carbon themselves or acquired it externally.


\section{Comparison to the Milky Way Stellar Halo}
\label{sec:mw}

The MW stellar halo most likely grows by the minor merging of
satellite galaxies.  As a result, the chemical abundance distribution
of the halo is the superposition of the abundance distributions of all
of its constituents.  The abundances of these individual satellites
depend on their stellar masses and SFHs.  Therefore, the mass
distribution and SFH of the halo's dissolved satellites can be
inferred from the chemical abundances of the halo
\citep[see][]{dlee14}.

One straightforward way to compare the carbon abundance distributions
of the surviving dSphs to the MW halo is to measure the frequency of
CEMP stars.  The frequency depends on metallicity.  Specifically, it
decreases toward higher metallicity \citep{bee92}.  In
Section~\ref{sec:dsphs}, we showed that [C/Fe] does indeed decrease
with increasing [Fe/H].  After accounting for this decrease with
metallicity, \citet{sta13} presented evidence that the CEMP fraction
in Sculptor is lower than in the MW halo with $\sim 90\%$ confidence.
This result would suggest that Sculptor had a different chemical
evolution or perhaps a different frequency of mass-transfer binaries
than halo progenitors.

Very few of the stars in our sample would qualify as CEMP under the
various definitions for ``CEMP.''  One proposed definition is a simple
cut in carbon abundances, such as $\cfe > +1.0$ \citep{bee05}.
However, this definition disregards the destruction of carbon on the
upper RGB, which is the location for all of the stars in our sample.
A more nuanced definition accounts for the luminosity of the star
\citep{aok07,yon13,pla14}.  We consider the CEMP fraction with two
different definitions.  First, we consider \citeauthor{aok07}'s
(\citeyear{aok07}) definition:

\begin{eqnarray}
\begin{array}{ll}
\cfe \ge +0.7 & ~~~{\rm if}~\log (L/L_{\sun}) \le 2.3 \\
\cfe \ge +3.0 - \log (L/L_{\sun}) & ~~~{\rm if}~\log (L/L_{\sun}) > 2.3 \\ \label{eq:aok07}
\end{array}
\end{eqnarray}

\noindent
The dotted line in Figure~\ref{fig:dsphcfelogl} shows the dividing
line of Equation~\ref{eq:aok07}.

We identified only \ncemp\ CEMP stars out of \ntot\ stars in our dSph
sample using \citeauthor{aok07}'s (\citeyear{aok07}) definition of
CEMP\@.  Of these, \ncb\ has $\feh \le \fehb$, \ncc\ have $\feh \le
\fehc$, and \ncf\ have $\feh \le \fehf$.  In these same metallicity
ranges, the CEMP fractions are $(\cempfracb \pm \cempfracerrb)\%$,
$(\cempfracc \pm \cempfracerrc)\%$, and $(\cempfracf \pm
\cempfracerrf)\%$.  However, our sample is heavily biased against
extremely carbon-rich stars.  First, we selected against stars redward
of the RGB, where C stars are generally found.  Second, our technique
of measuring [Fe/H] in the range 6300--9100~\AA\ fails for extremely
carbon-rich stars, in which CN blankets much of that spectral region.
We found \ncstar\ C stars, but we did not measure their metallicities.
As a result, they are not included in the above estimate of CEMP
frequency.  Therefore, the frequencies of carbon stars we found in
dSphs should be regarded strictly as lower limits.

We also counted CEMP stars using $\cfecorr$, which is corrected for
luminosity according to \citet{pla14}.  One of \citeauthor{pla14}'s
proposed definitions for CEMP is $\cfecorr > +0.7$ (dotted line in
Figure~\ref{fig:dsphcfefeh}).  The cut does not need any dependence on
luminosity because $\cfecorr$ already depends on luminosity.  Under
this definition, we found only \ncemppla\ out of \ntot\ stars.  They
have $\feh = \cempplafehc \pm \cempplafeherrc$, $\cempplafeha \pm
\cempplafeherra$, and $\cempplafehb \pm \cempplafeherrb$.

The CEMP fraction of the halo is significantly higher than the lower
limits we determined for dSphs except in the $\feh \le \fehb$ range,
where we observed only \ntb\ stars.  In the halo, the CEMP fraction is
approximately 13\% at $\feh \le -2$ and 25\% at $\feh \le -3$
\citep{coh05,mar05,luc06,fre06,car12,yon13,lee13,pla14}.  Many of the
halo's CEMP stars have $\cfe \ga +2$.  These stars would have been
excluded from our sample.

\begin{figure}[t!]
\centering
\includegraphics[width=\columnwidth]{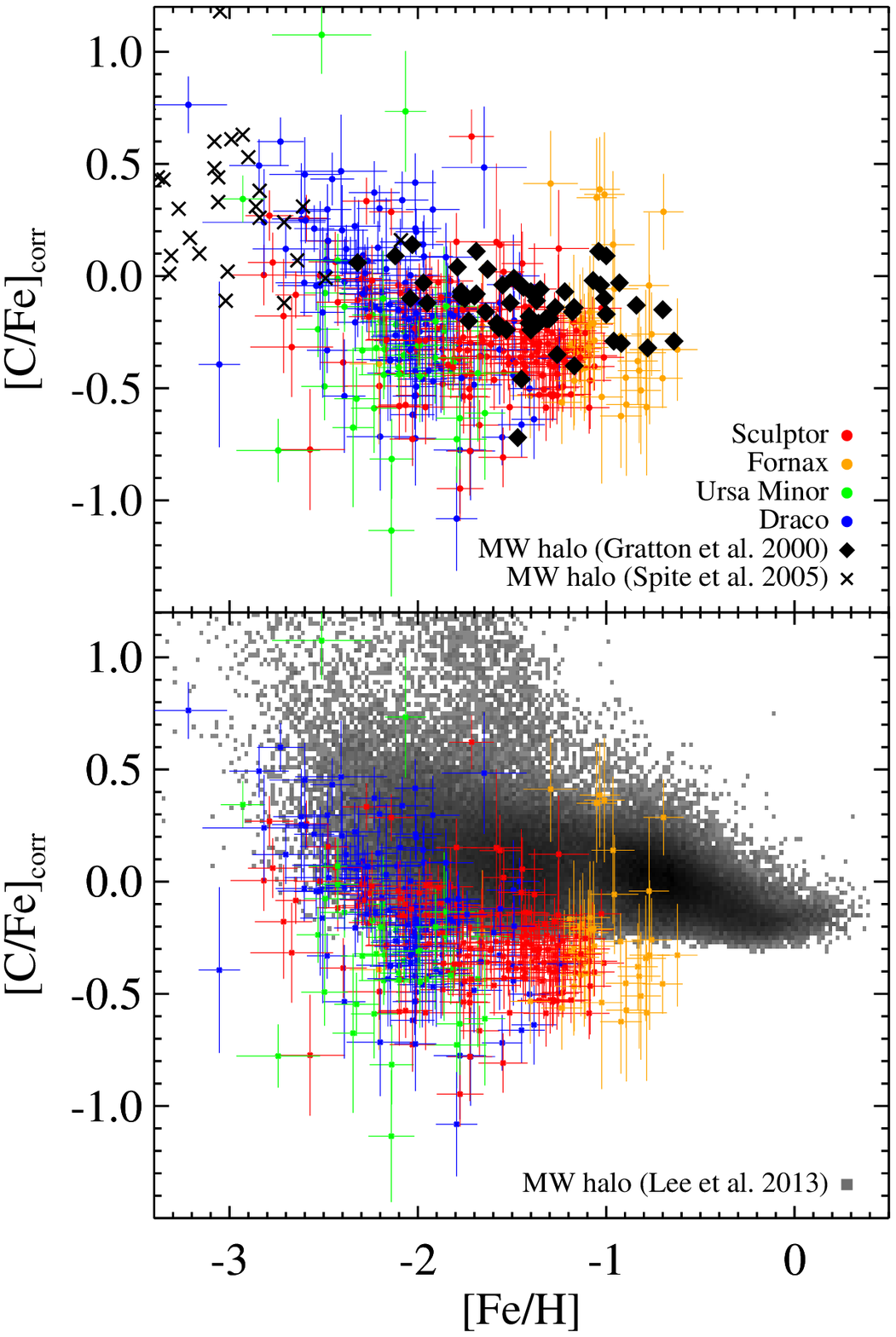}
\caption{The trend of $\cfecorr$ versus [Fe/H] in dSphs compared to
  stars in the MW (top panel: \citealt{gra00} and \citealt{spi05};
  bottom panel: \citealt{lee13}).  In the bottom panel, the grayscale
  for MW stars is logarithmic in number density.  The bottom panel
  shows only main sequence turn-off stars, which were selected in the
  same manner as in \citeauthor{lee13}: $5600~{\rm K} \le T_{\rm eff}
  \le 6700~{\rm K}$.  These stars have not altered their surface
  compositions, so their [C/Fe] measurements do not need to be
  corrected.\label{fig:halo}}
\end{figure}

Nonetheless, we can still compare the general trend of [C/Fe] versus
[Fe/H] for stars with $\cfe < +1$ between dSphs and the halo.
Although our sample misses many of the stars with $\cfe \ga +1$, the
remaining stars show a trend different from the MW halo, as shown in
Figure~\ref{fig:halo}.  The top panel compares our measurements of
$\cfecorr$ in dSphs compared to high-resolution measurements in the MW
halo \citep{gra00,spi05}.  The sample of \citeauthor{gra00}\ includes
dwarfs and giants, and the sample of \citeauthor{spi05}\ is only
giants.  We applied \citeauthor{pla14}'s (\citeyear{pla14})
corrections to the [C/Fe] measurements in both samples.  The bottom
panel compares our measurements to SEGUE main sequence turn-off stars
\citep{lee13}, measured by the Sloan Extension for Galactic
Understanding and Exploration \citep[SEGUE,][]{yan09}.  We plot the
turn-off stars because their carbon abundances have not yet been
affected by stellar astration.  We adjusted all of the MW [C/Fe]
measurements to match the solar abundance ratios we adopted (see
Section~\ref{sec:validation}).  Ignoring the CEMP stars, the [C/Fe]
pattern of the general halo population is flat at $\cfe \sim +0.1$
until $\feh = -0.7$, whereupon it drops to $\cfe \sim -0.2$ at $\feh =
0$.  In contrast, the dSph stars have already dropped to $\cfe \sim
-0.4$ by a metallicity of $\feh = -2$.  Note that the floor in the
SEGUE data at $\cfe \sim -0.3$ might be artificial.  The lowest value
of [C/Fe] in \citeauthor{lee13}'s grid of synthetic spectra is $-0.5$,
and the minimum upper limit values (not plotted in
Figure~\ref{fig:halo}) are mostly about 0.2~dex away from the grid
limit.  Regardless, the SEGUE sample is qualitatively consistent with
the samples of \citet{gra00} and \citet{spi05}.

\citet{lee14} interpreted the halo's CEMP fraction in the context of
binary mass transfer.  They found that carbon donation from AGB stars
could explain the present [C/Fe] distribution for giant stars.
However, that mechanism is primarily responsible for large carbon
enhancements ($\cfe > +1$).  In Section~\ref{sec:dsphs} and
Figure~\ref{fig:nucleosynthesis}, we interpreted the trend among
non-CEMP stars in the context of galactic chemical evolution.  The
different [C/Fe] slopes between the dSph and halo samples is
reminiscent of the different slopes in [$\alpha$/Fe]
\citep[e.g.,][]{she98a,ven04}.  The [$\alpha$/Fe] distribution of the
halo shows a downturn (a ``knee'') at $\feh = -0.7$, a feature that is
also visible in [C/Fe] (Figure~\ref{fig:halo}).  The knee corresponds
to the metallicity at which Type~Ia SNe began to pollute the MW\@.
Because Type~Ia SNe have low yields of both [$\alpha$/Fe] and [C/Fe],
their prevalence beginning at $\feh = -0.7$ can explain the abundance
pattern for both abundance ratios in the halo.  However, the knee
happens at lower metallicities for dSphs.  The decline of [C/Fe] with
increasing [Fe/H] begins at $\feh \la -2.5$ for Ursa Minor and Draco.
The lower metallicity of their [C/Fe] knees indicates that Type~Ia SNe
were active at much lower metallicities in dSphs than in the halo.
The same trend is seen in dSphs' [$\alpha$/Fe] ratios
\citep[e.g.,][]{ven04,kir11b}.  The halo reached a much higher
metallicity than the dSphs before Type~Ia SNe turned on.  Furthermore,
both [C/Fe] and [$\alpha$/Fe] in dSphs decline to values well below
the solar ratio, indicating a heavy dose of Type~Ia SNe relative to
the halo, which has larger values of [C/Fe] and [$\alpha$/Fe].  Our
[C/Fe] measurements are further proof that the chemical evolution of
surviving dSphs was much slower than the progenitors of the MW halo.


\section{Summary}
\label{sec:summary}

Carbon is a complex element.  It has diverse nucleosynthetic sources.
AGB stars produce copious amounts of carbon, but their C/Fe ratios
depend sensitively on metallicity and stellar mass.  Type~II SNe
produce some carbon, whereas Type~Ia SNe produce very little carbon
but a great deal of iron.  The timescale for all three sources is
different, which makes the C/Fe ratio change as the galaxy evolves in
a way that depends on the details of the SFH\@.  Mass transfer from
past AGB stars further complicates the interpretation of the carbon
abundances of present-day stars.  In particular, some stars in
mass-transfer binaries can have orders of magnitude more carbon than
single stars of the same metallicity.  In addition to multiple
sources, carbon also has a sink.  Red giants astrate carbon on the
upper RGB\@.  The rate of astration depends on the metallicity and
luminosity of the star.

In order to explore the evolution of carbon, we observed
\ndsphgctot\ red giants with the Keck/DEIMOS spectrograph.  We
generated a grid of model spectra based on a line list that we checked
against spectra of the Sun and Arcturus.  We measured carbon
abundances by fitting the model spectra to the G band at 4300~\AA\@.
We were able to measure [C/Fe] for \ntotch\ stars.  We validated our
measurements by comparing a subset of our observations to published
measurements of carbon abundances for the same stars.  We found
offsets between our measurements of [C/Fe] and the literature
measurements.  However, the offset depended on the source of the
measurements.  Our measurements agreed most closely with those based
on atomic and molecular line lists most similar to our own.

Globular clusters provide controlled environments to study the
evolution of carbon because they generally have single ages and
metallicities.  We measured [C/Fe] in \ngctot\ red giants in the GCs
NGC~2419, M68, and M15.  We observed the well-known decline of [C/Fe]
on the upper RGB starting at the luminosity function bump.  We found
that carbon does not participate in the unusual abundance patterns in
NGC~2419 \citep{coh12,muc12}.  We also reported the discovery of a
carbon-enhanced AGB star (\gccrichname) in M15.

We also explored the evolution of carbon in MW dSphs by examining
\ndsphch\ carbon abundances and \ndsphul\ upper limits for stars
confirmed to be members of Sculptor, Fornax, Ursa Minor, and Draco.
As expected, the [C/Fe] distribution is more complex in dSphs than in
GCs.  Unlike GCs, dSphs show a dispersion in [C/Fe] at fixed stellar
luminosity.  The dispersion depends on [Fe/H].  Specifically, stars
with lower [Fe/H] have higher [C/Fe] than stars at the same luminosity
with higher [Fe/H].  We interpreted the trend with metallicity as an
evolution in the influence of the three sources of carbon.  First,
Type~II SNe seeded the dSph with the first C and Fe.  Very soon after,
the first AGB stars produced a great deal of C relative to Fe.
Finally, Type~Ia SNe produced very little C but a large amount of Fe
after $\sim 100$~Myr.  The result is a decline in [C/Fe] with
increasing [Fe/H] that roughly tracks the decline in [$\alpha$/Fe],
also due to Type~Ia SNe \citep[e.g.,][]{kir11b}.

Our color selection biased the sample against very carbon-rich stars.
Nonetheless, we found \ncstar\ C stars (8 previously known) in three
dSphs.  These stars are either AGB stars that are producing carbon now
or the results of binary mass transfer from past AGB stars.  Because
of our selection bias, we calculated only lower limits to the CEMP
fraction in dSphs.  These lower limits are mostly below the CEMP
fraction for the MW halo.  Therefore, we have found no tension for the
theory of hierarchical assembly, wherein the halo builds mass through
minor merging of satellites like dSphs \citep[e.g.,][]{bul05}.
However, this question would be addressed better by a spectroscopic
survey of dSphs with a color cut more inclusive of stars redder than
the RGB\@.

Our sample is still useful for comparing the trend of [C/Fe] versus
[Fe/H] for carbon-normal stars.  We found that the knee in the [C/Fe]
distribution occurs at a lower metallicity in dSphs than in the MW
halo \citep{lee13}.  This difference in metallicity is reminiscent of
the different [$\alpha$/Fe] knees between dSphs and the MW halo
\citep{ven04}.  The similarity in behavior between [C/Fe] and
      [$\alpha$/Fe] supports our hypothesis that Type~Ia SNe are
      responsible for the decline in [C/Fe].  The [$\alpha$/Fe] and
      [C/Fe] distributions together indicate that the progenitors of
      the MW halo had a more efficient SFH, which reached a higher
      metallicity before Type~Ia SNe began to produce Fe in earnest.

Our sample of \ntot\ measurements doubles the number of carbon
abundances known in dSphs.  Each dSph has enough carbon measurements
(\nscl\ in Sculptor, \nfor\ in Fornax, \numi\ in Ursa Minor, and
\ndra\ in Draco) for analysis with a numerical chemical evolution
model \citep[e.g.,][]{rev09,kir11b}.  Although we have presented a
qualitative interpretation for the observed carbon abundance trends, a
numerical model will give a quantitative timescale for star formation.
This chemical timescale can be compared to photometric measurements of
SFH\@.  Furthermore, numerical models can possibly constrain the IMF
\citep{lee14} or binary fraction \citep{luc05} in dSphs.  We invite
interested modelers to make quantitative models of the carbon
abundances in Table~\ref{tab:cfe} to reveal the history of star
formation in these four dSphs.

\acknowledgments We thank Vinicius Placco for kindly computing
astration corrections to carbon abundances.  We also thank Chris
Sneden for constructive comments on the structure of this paper and
the anonymous referee for an insightful report.  M.G., A.J.Z., and
M.D.\ carried out their work through UCSC's Science Internship Program
for high-school students.  P.G.\ acknowledges support from NSF grants
AST-1010039 and AST-1412648.

We are grateful to the many people who have worked to make the Keck
Telescope and its instruments a reality and to operate and maintain
the Keck Observatory.  The authors wish to extend special thanks to
those of Hawaiian ancestry on whose sacred mountain we are privileged
to be guests.  Without their generous hospitality, none of the
observations presented herein would have been possible.

{\it Facility:} \facility{Keck:II (DEIMOS)}

\bibliography{Kirby_et_al_carbon}

\begin{thebibliography}{}
\expandafter\ifx\csname natexlab\endcsname\relax\def\natexlab#1{#1}\fi

\bibitem[{{Aaronson} {et~al.}(1983){Aaronson}, {Hodge}, \& {Olszewski}}]{aar83}
{Aaronson}, M., {Hodge}, P.~W., \& {Olszewski}, E.~W. 1983, \apj, 267, 271

\bibitem[{{Aaronson} {et~al.}(1982){Aaronson}, {Liebert}, \& {Stocke}}]{aar82}
{Aaronson}, M., {Liebert}, J., \& {Stocke}, J. 1982, \apj, 254, 507

\bibitem[{{Aaronson} \& {Mould}(1980)}]{aar80}
{Aaronson}, M., \& {Mould}, J. 1980, \apj, 240, 804

\bibitem[{{Abia} {et~al.}(2008){Abia}, {de Laverny}, \& {Wahlin}}]{abi08}
{Abia}, C., {de Laverny}, P., \& {Wahlin}, R. 2008, \aap, 481, 161

\bibitem[{{Alcaino} {et~al.}(1990){Alcaino}, {Liller}, {Alvarado}, \&
  {Wenderoth}}]{alc90}
{Alcaino}, G., {Liller}, W., {Alvarado}, F., \& {Wenderoth}, E. 1990, \aj, 99,
  1831

\bibitem[{{Anders} \& {Grevesse}(1989)}]{and89}
{Anders}, E., \& {Grevesse}, N. 1989, \gca, 53, 197

\bibitem[{{Aoki} {et~al.}(2007){Aoki}, {Beers}, {Christlieb}, {Norris}, {Ryan},
  \& {Tsangarides}}]{aok07}
{Aoki}, W., {Beers}, T.~C., {Christlieb}, N., {et~al.} 2007, \apj, 655, 492

\bibitem[{{Armandroff} {et~al.}(1995){Armandroff}, {Olszewski}, \&
  {Pryor}}]{arm95}
{Armandroff}, T.~E., {Olszewski}, E.~W., \& {Pryor}, C. 1995, \aj, 110, 2131

\bibitem[{{Azzopardi} {et~al.}(1986){Azzopardi}, {Lequeux}, \&
  {Westerlund}}]{azz86}
{Azzopardi}, M., {Lequeux}, J., \& {Westerlund}, B.~E. 1986, \aap, 161, 232

\bibitem[{{Beers} \& {Christlieb}(2005)}]{bee05}
{Beers}, T.~C., \& {Christlieb}, N. 2005, \araa, 43, 531

\bibitem[{{Beers} {et~al.}(1992){Beers}, {Preston}, \& {Shectman}}]{bee92}
{Beers}, T.~C., {Preston}, G.~W., \& {Shectman}, S.~A. 1992, \aj, 103, 1987

\bibitem[{{Behr} {et~al.}(1999){Behr}, {Cohen}, {McCarthy}, \&
  {Djorgovski}}]{beh99}
{Behr}, B.~B., {Cohen}, J.~G., {McCarthy}, J.~K., \& {Djorgovski}, S.~G. 1999,
  \apjl, 517, L135

\bibitem[{{Bell}(1985)}]{bel85}
{Bell}, R.~A. 1985, \pasp, 97, 219

\bibitem[{{Bellazzini} {et~al.}(2002){Bellazzini}, {Ferraro}, {Origlia},
  {Pancino}, {Monaco}, \& {Oliva}}]{bel02}
{Bellazzini}, M., {Ferraro}, F.~R., {Origlia}, L., {et~al.} 2002, \aj, 124,
  3222

\bibitem[{{B{\"o}hm-Vitense} {et~al.}(2000){B{\"o}hm-Vitense}, {Carpenter},
  {Robinson}, {Ake}, \& {Brown}}]{boh00}
{B{\"o}hm-Vitense}, E., {Carpenter}, K., {Robinson}, R., {Ake}, T., \& {Brown},
  J. 2000, \apj, 533, 969

\bibitem[{{Bond}(1975)}]{bon75}
{Bond}, H.~E. 1975, \apjl, 202, L47

\bibitem[{{Bond} \& {Neff}(1969)}]{bon69}
{Bond}, H.~E., \& {Neff}, J.~S. 1969, \apj, 158, 1235

\bibitem[{{Briley} {et~al.}(2002){Briley}, {Cohen}, \& {Stetson}}]{bri02}
{Briley}, M.~M., {Cohen}, J.~G., \& {Stetson}, P.~B. 2002, \apjl, 579, L17

\bibitem[{{Bullock} \& {Johnston}(2005)}]{bul05}
{Bullock}, J.~S., \& {Johnston}, K.~V. 2005, \apj, 635, 931

\bibitem[{{Canterna} \& {Schommer}(1978)}]{can78}
{Canterna}, R., \& {Schommer}, R.~A. 1978, \apjl, 219, L119

\bibitem[{{Carbon} {et~al.}(1982){Carbon}, {Romanishin}, {Langer}, {Butler},
  {Kemper}, {Trefzger}, {Kraft}, \& {Suntzeff}}]{car82}
{Carbon}, D.~F., {Romanishin}, W., {Langer}, G.~E., {et~al.} 1982, \apjs, 49,
  207

\bibitem[{{Carollo} {et~al.}(2012){Carollo}, {Beers}, {Bovy}, {Sivarani},
  {Norris}, {Freeman}, {Aoki}, {Lee}, \& {Kennedy}}]{car12}
{Carollo}, D., {Beers}, T.~C., {Bovy}, J., {et~al.} 2012, \apj, 744, 195

\bibitem[{{Charbonnel} \& {Lagarde}(2010)}]{cha10}
{Charbonnel}, C., \& {Lagarde}, N. 2010, \aap, 522, A10

\bibitem[{{Charbonnel} \& {Zahn}(2007)}]{cha07}
{Charbonnel}, C., \& {Zahn}, J.-P. 2007, \aap, 467, L15

\bibitem[{{Cohen}(1999)}]{coh99b}
{Cohen}, J.~G. 1999, \aj, 117, 2434

\bibitem[{{Cohen} \& {Huang}(2009)}]{coh09}
{Cohen}, J.~G., \& {Huang}, W. 2009, \apj, 701, 1053

\bibitem[{{Cohen} \& {Huang}(2010)}]{coh10}
---. 2010, \apj, 719, 931

\bibitem[{{Cohen} {et~al.}(2011){Cohen}, {Huang}, \& {Kirby}}]{coh11}
{Cohen}, J.~G., {Huang}, W., \& {Kirby}, E.~N. 2011, \apj, 740, 60

\bibitem[{{Cohen} \& {Kirby}(2012)}]{coh12}
{Cohen}, J.~G., \& {Kirby}, E.~N. 2012, \apj, 760, 86

\bibitem[{{Cohen} {et~al.}(1978){Cohen}, {Persson}, \& {Frogel}}]{coh78}
{Cohen}, J.~G., {Persson}, S.~E., \& {Frogel}, J.~A. 1978, \apj, 222, 165

\bibitem[{{Cohen} {et~al.}(2005){Cohen}, {Shectman}, {Thompson}, {McWilliam},
  {Christlieb}, {Melendez}, {Zickgraf}, {Ram{\'{\i}}rez}, \& {Swenson}}]{coh05}
{Cohen}, J.~G., {Shectman}, S., {Thompson}, I., {et~al.} 2005, \apjl, 633, L109

\bibitem[{{Cohen} {et~al.}(2006){Cohen}, {McWilliam}, {Shectman}, {Thompson},
  {Christlieb}, {Melendez}, {Ramirez}, {Swensson}, \& {Zickgraf}}]{coh06}
{Cohen}, J.~G., {McWilliam}, A., {Shectman}, S., {et~al.} 2006, \aj, 132, 137

\bibitem[{{Cooper} {et~al.}(2012){Cooper}, {Newman}, {Davis}, {Finkbeiner}, \&
  {Gerke}}]{coo12}
{Cooper}, M.~C., {Newman}, J.~A., {Davis}, M., {Finkbeiner}, D.~P., \& {Gerke},
  B.~F. 2012, {spec2d: DEEP2 DEIMOS Spectral Pipeline}, astrophysics Source
  Code Library, ascl:1203.003

\bibitem[{{C{\^o}t{\'e}} {et~al.}(1997){C{\^o}t{\'e}}, {Hanes}, {McLaughlin},
  {Bridges}, {Hesser}, \& {Harris}}]{cot97}
{C{\^o}t{\'e}}, P., {Hanes}, D.~A., {McLaughlin}, D.~E., {et~al.} 1997, \apjl,
  476, L15

\bibitem[{{Cowley} \& {Crampton}(1985)}]{cow85}
{Cowley}, A.~P., \& {Crampton}, D. 1985, \pasp, 97, 835

\bibitem[{{Dahn} {et~al.}(1977){Dahn}, {Liebert}, {Kron}, {Spinrad}, \&
  {Hintzen}}]{dah77}
{Dahn}, C.~C., {Liebert}, J., {Kron}, R.~G., {Spinrad}, H., \& {Hintzen}, P.~M.
  1977, \apj, 216, 757

\bibitem[{{de Boer} {et~al.}(2012){de Boer}, {Tolstoy}, {Hill}, {Saha},
  {Olszewski}, {Mateo}, {Starkenburg}, {Battaglia}, \& {Walker}}]{deb12}
{de Boer}, T.~J.~L., {Tolstoy}, E., {Hill}, V., {et~al.} 2012, \aap, 544, A73

\bibitem[{{Demarque} {et~al.}(2004){Demarque}, {Woo}, {Kim}, \& {Yi}}]{dem04}
{Demarque}, P., {Woo}, J.-H., {Kim}, Y.-C., \& {Yi}, S.~K. 2004, \apjs, 155,
  667

\bibitem[{{Demers} \& {Battinelli}(2002)}]{dem02}
{Demers}, S., \& {Battinelli}, P. 2002, \aj, 123, 238

\bibitem[{{Dickens}(1972)}]{dic72}
{Dickens}, R.~J. 1972, \mnras, 159, 7P

\bibitem[{{Durrell} \& {Harris}(1993)}]{dur93}
{Durrell}, P.~R., \& {Harris}, W.~E. 1993, \aj, 105, 1420

\bibitem[{{Eggleton} {et~al.}(2006){Eggleton}, {Dearborn}, \&
  {Lattanzio}}]{egg06}
{Eggleton}, P.~P., {Dearborn}, D.~S.~P., \& {Lattanzio}, J.~C. 2006, Science,
  314, 1580

\bibitem[{{Faber} {et~al.}(2003){Faber}, {Phillips}, {Kibrick}, {Alcott},
  {Allen}, {Burrous}, {Cantrall}, {Clarke}, {Coil}, {Cowley}, {Davis}, {Deich},
  {Dietsch}, {Gilmore}, {Harper}, {Hilyard}, {Lewis}, {McVeigh}, {Newman},
  {Osborne}, {Schiavon}, {Stover}, {Tucker}, {Wallace}, {Wei}, {Wirth}, \&
  {Wright}}]{fab03}
{Faber}, S.~M., {Phillips}, A.~C., {Kibrick}, R.~I., {et~al.} 2003, in Society
  of Photo-Optical Instrumentation Engineers (SPIE) Conference Series, Vol.
  4841, Instrument Design and Performance for Optical/Infrared Ground-based
  Telescopes, ed. M.~{Iye} \& A.~F.~M. {Moorwood}, 1657--1669

\bibitem[{{Feast} {et~al.}(2013){Feast}, {Menzies}, \& {Whitelock}}]{fea13}
{Feast}, M.~W., {Menzies}, J.~W., \& {Whitelock}, P.~A. 2013, \mnras, 428, L36

\bibitem[{{Font} {et~al.}(2006){Font}, {Johnston}, {Bullock}, \&
  {Robertson}}]{fon06}
{Font}, A.~S., {Johnston}, K.~V., {Bullock}, J.~S., \& {Robertson}, B.~E. 2006,
  \apj, 638, 585

\bibitem[{{Frebel} {et~al.}(2010{\natexlab{a}}){Frebel}, {Kirby}, \&
  {Simon}}]{fre10a}
{Frebel}, A., {Kirby}, E.~N., \& {Simon}, J.~D. 2010{\natexlab{a}}, \nat, 464,
  72

\bibitem[{{Frebel} {et~al.}(2010{\natexlab{b}}){Frebel}, {Simon}, {Geha}, \&
  {Willman}}]{fre10b}
{Frebel}, A., {Simon}, J.~D., {Geha}, M., \& {Willman}, B. 2010{\natexlab{b}},
  \apj, 708, 560

\bibitem[{{Frebel} {et~al.}(2014){Frebel}, {Simon}, \& {Kirby}}]{fre14}
{Frebel}, A., {Simon}, J.~D., \& {Kirby}, E.~N. 2014, \apj, 786, 74

\bibitem[{{Frebel} {et~al.}(2006){Frebel}, {Christlieb}, {Norris}, {Beers},
  {Bessell}, {Rhee}, {Fechner}, {Marsteller}, {Rossi}, {Thom}, {Wisotzki}, \&
  {Reimers}}]{fre06}
{Frebel}, A., {Christlieb}, N., {Norris}, J.~E., {et~al.} 2006, \apj, 652, 1585

\bibitem[{{Fulbright} {et~al.}(2004){Fulbright}, {Rich}, \& {Castro}}]{ful04}
{Fulbright}, J.~P., {Rich}, R.~M., \& {Castro}, S. 2004, \apj, 612, 447

\bibitem[{{Gilmore} \& {Wyse}(1991)}]{gil91}
{Gilmore}, G., \& {Wyse}, R.~F.~G. 1991, \apjl, 367, L55

\bibitem[{{Girardi} {et~al.}(2004){Girardi}, {Grebel}, {Odenkirchen}, \&
  {Chiosi}}]{gir04}
{Girardi}, L., {Grebel}, E.~K., {Odenkirchen}, M., \& {Chiosi}, C. 2004, \aap,
  422, 205

\bibitem[{{Gratton} {et~al.}(2004){Gratton}, {Sneden}, \& {Carretta}}]{gra04}
{Gratton}, R., {Sneden}, C., \& {Carretta}, E. 2004, \araa, 42, 385

\bibitem[{{Gratton} {et~al.}(2000){Gratton}, {Sneden}, {Carretta}, \&
  {Bragaglia}}]{gra00}
{Gratton}, R.~G., {Sneden}, C., {Carretta}, E., \& {Bragaglia}, A. 2000, \aap,
  354, 169

\bibitem[{{Harbeck} {et~al.}(2003){Harbeck}, {Smith}, \& {Grebel}}]{har03}
{Harbeck}, D., {Smith}, G.~H., \& {Grebel}, E.~K. 2003, \aj, 125, 197

\bibitem[{{Harding}(1962)}]{har62}
{Harding}, G.~A. 1962, The Observatory, 82, 205

\bibitem[{{Harris}(1996)}]{har96}
{Harris}, W.~E. 1996, \aj, 112, 1487

\bibitem[{{Harris} {et~al.}(1997){Harris}, {Bell}, {Vandenberg}, {Bolte},
  {Stetson}, {Hesser}, {van den Bergh}, {Bond}, {Fahlman}, \& {Richer}}]{har97}
{Harris}, W.~E., {Bell}, R.~A., {Vandenberg}, D.~A., {et~al.} 1997, \aj, 114,
  1030

\bibitem[{{Hilker} {et~al.}(2004){Hilker}, {Kayser}, {Richtler}, \&
  {Willemsen}}]{hil04}
{Hilker}, M., {Kayser}, A., {Richtler}, T., \& {Willemsen}, P. 2004, \aap, 422,
  L9

\bibitem[{{Hinkle} {et~al.}(2000){Hinkle}, {Wallace}, {Valenti}, \&
  {Harmer}}]{hin00}
{Hinkle}, K., {Wallace}, L., {Valenti}, J., \& {Harmer}, D. 2000, {Visible and
  Near Infrared Atlas of the Arcturus Spectrum 3727-9300 A}

\bibitem[{{Iben}(1975)}]{ibe75}
{Iben}, Jr., I. 1975, \apj, 196, 525

\bibitem[{{Iwamoto} {et~al.}(1999){Iwamoto}, {Brachwitz}, {Nomoto},
  {Kishimoto}, {Umeda}, {Hix}, \& {Thielemann}}]{iwa99}
{Iwamoto}, K., {Brachwitz}, F., {Nomoto}, K., {et~al.} 1999, \apjs, 125, 439

\bibitem[{{Jorgensen} {et~al.}(1996){Jorgensen}, {Larsson}, {Iwamae}, \&
  {Yu}}]{jor96}
{Jorgensen}, U.~G., {Larsson}, M., {Iwamae}, A., \& {Yu}, B. 1996, \aap, 315,
  204

\bibitem[{{Karakas}(2010)}]{kar10}
{Karakas}, A.~I. 2010, \mnras, 403, 1413

\bibitem[{{Karakas} {et~al.}(2012){Karakas}, {Garc{\'{\i}}a-Hern{\'a}ndez}, \&
  {Lugaro}}]{kar12}
{Karakas}, A.~I., {Garc{\'{\i}}a-Hern{\'a}ndez}, D.~A., \& {Lugaro}, M. 2012,
  \apj, 751, 8

\bibitem[{{Karinkuzhi} \& {Goswami}(2014)}]{kar14}
{Karinkuzhi}, D., \& {Goswami}, A. 2014, \mnras, 440, 1095

\bibitem[{{Keller} {et~al.}(2001){Keller}, {Pilachowski}, \& {Sneden}}]{kel01}
{Keller}, L.~D., {Pilachowski}, C.~A., \& {Sneden}, C. 2001, \aj, 122, 2554

\bibitem[{{Kinman} {et~al.}(1981){Kinman}, {Carbon}, {Suntzeff}, \&
  {Kraft}}]{kin81}
{Kinman}, T.~D., {Carbon}, D.~F., {Suntzeff}, N., \& {Kraft}, R.~P. 1981, in
  IAU Colloq. 68: Astrophysical Parameters for Globular Clusters, ed. A.~G.~D.
  {Philip} \& D.~S. {Hayes}, 451

\bibitem[{{Kirby}(2011)}]{kir11}
{Kirby}, E.~N. 2011, \pasp, 123, 531

\bibitem[{{Kirby} {et~al.}(2011){Kirby}, {Cohen}, {Smith}, {Majewski}, {Sohn},
  \& {Guhathakurta}}]{kir11b}
{Kirby}, E.~N., {Cohen}, J.~G., {Smith}, G.~H., {et~al.} 2011, \apj, 727, 79

\bibitem[{{Kirby} {et~al.}(2009){Kirby}, {Guhathakurta}, {Bolte}, {Sneden}, \&
  {Geha}}]{kir09}
{Kirby}, E.~N., {Guhathakurta}, P., {Bolte}, M., {Sneden}, C., \& {Geha}, M.~C.
  2009, \apj, 705, 328

\bibitem[{{Kirby} {et~al.}(2008){Kirby}, {Guhathakurta}, \& {Sneden}}]{kir08}
{Kirby}, E.~N., {Guhathakurta}, P., \& {Sneden}, C. 2008, \apj, 682, 1217

\bibitem[{{Kirby} {et~al.}(2010){Kirby}, {Guhathakurta}, {Simon}, {Geha},
  {Rockosi}, {Sneden}, {Cohen}, {Sohn}, {Majewski}, \& {Siegel}}]{kir10}
{Kirby}, E.~N., {Guhathakurta}, P., {Simon}, J.~D., {et~al.} 2010, \apjs, 191,
  352

\bibitem[{{Koch} {et~al.}(2013){Koch}, {Feltzing}, {Ad{\'e}n}, \&
  {Matteucci}}]{koc13}
{Koch}, A., {Feltzing}, S., {Ad{\'e}n}, D., \& {Matteucci}, F. 2013, \aap, 554,
  A5

\bibitem[{{Koch} \& {Rich}(2014)}]{koc14}
{Koch}, A., \& {Rich}, R.~M. 2014, \apj, 794, 89

\bibitem[{Kramida {et~al.}(2014)Kramida, {Yu.~Ralchenko}, Reader, \& {and NIST
  ASD Team}}]{kra14}
Kramida, A., {Yu.~Ralchenko}, Reader, J., \& {and NIST ASD Team}. 2014, {NIST
  Atomic Spectra Database (ver. 5.2), [Online]. Available:
  {\tt{http://physics.nist.gov/asd}} [2014, October 17]. National Institute of
  Standards and Technology, Gaithersburg, MD.}

\bibitem[{{Kupka} {et~al.}(1999){Kupka}, {Piskunov}, {Ryabchikova}, {Stempels},
  \& {Weiss}}]{kup99}
{Kupka}, F., {Piskunov}, N., {Ryabchikova}, T.~A., {Stempels}, H.~C., \&
  {Weiss}, W.~W. 1999, \aaps, 138, 119

\bibitem[{{Kurucz}(1993)}]{kur93}
{Kurucz}, R. 1993, ATLAS9 Stellar Atmosphere Programs and 2 km/s grid.~Kurucz
  CD-ROM No.~13.~ Cambridge, Mass.: Smithsonian Astrophysical Observatory,
  1993., 13

\bibitem[{{Kurucz}(1992)}]{kur92}
{Kurucz}, R.~L. 1992, Revista Mexicana de Astronomia y Astrofisica, 23, 45

\bibitem[{{Lai} {et~al.}(2011){Lai}, {Lee}, {Bolte}, {Lucatello}, {Beers},
  {Johnson}, {Sivarani}, \& {Rockosi}}]{lai11}
{Lai}, D.~K., {Lee}, Y.~S., {Bolte}, M., {et~al.} 2011, \apj, 738, 51

\bibitem[{{Lee} {et~al.}(2014{\natexlab{a}}){Lee}, {Johnston}, {Sen}, \&
  {Jessop}}]{dlee14}
{Lee}, D.~M., {Johnston}, K.~V., {Sen}, B., \& {Jessop}, W. 2014{\natexlab{a}},
  \apj, in press, arXiv:1410.6166

\bibitem[{{Lee} {et~al.}(2005){Lee}, {Carney}, \& {Habgood}}]{lee05}
{Lee}, J.-W., {Carney}, B.~W., \& {Habgood}, M.~J. 2005, \aj, 129, 251

\bibitem[{{Lee} {et~al.}(2014{\natexlab{b}}){Lee}, {Suda}, {Beers}, \&
  {Stancliffe}}]{lee14}
{Lee}, Y.~S., {Suda}, T., {Beers}, T.~C., \& {Stancliffe}, R.~J.
  2014{\natexlab{b}}, \apj, 788, 131

\bibitem[{{Lee} {et~al.}(2013){Lee}, {Beers}, {Masseron}, {Plez}, {Rockosi},
  {Sobeck}, {Yanny}, {Lucatello}, {Sivarani}, {Placco}, \& {Carollo}}]{lee13}
{Lee}, Y.~S., {Beers}, T.~C., {Masseron}, T., {et~al.} 2013, \aj, 146, 132

\bibitem[{{Liang} {et~al.}(2000){Liang}, {Zhao}, \& {Zhang}}]{lia00}
{Liang}, Y.~C., {Zhao}, G., \& {Zhang}, B. 2000, \aap, 363, 555

\bibitem[{{Lind} {et~al.}(2011){Lind}, {Charbonnel}, {Decressin}, {Primas},
  {Grundahl}, \& {Asplund}}]{lin11}
{Lind}, K., {Charbonnel}, C., {Decressin}, T., {et~al.} 2011, \aap, 527, A148

\bibitem[{{Lucatello} {et~al.}(2006){Lucatello}, {Beers}, {Christlieb},
  {Barklem}, {Rossi}, {Marsteller}, {Sivarani}, \& {Lee}}]{luc06}
{Lucatello}, S., {Beers}, T.~C., {Christlieb}, N., {et~al.} 2006, \apjl, 652,
  L37

\bibitem[{{Lucatello} {et~al.}(2005){Lucatello}, {Tsangarides}, {Beers},
  {Carretta}, {Gratton}, \& {Ryan}}]{luc05}
{Lucatello}, S., {Tsangarides}, S., {Beers}, T.~C., {et~al.} 2005, \apj, 625,
  825

\bibitem[{{Markwardt}(2012)}]{mar12}
{Markwardt}, C. 2012, {MPFIT: Robust non-linear least squares curve fitting},
  astrophysics Source Code Library, ascl:1208.019

\bibitem[{{Marsteller} {et~al.}(2005){Marsteller}, {Beers}, {Rossi},
  {Christlieb}, {Bessell}, \& {Rhee}}]{mar05}
{Marsteller}, B., {Beers}, T.~C., {Rossi}, S., {et~al.} 2005, Nuclear Physics
  A, 758, 312

\bibitem[{{Martell} {et~al.}(2008{\natexlab{a}}){Martell}, {Smith}, \&
  {Briley}}]{mar08a}
{Martell}, S.~L., {Smith}, G.~H., \& {Briley}, M.~M. 2008{\natexlab{a}}, \pasp,
  120, 839

\bibitem[{{Martell} {et~al.}(2008{\natexlab{b}}){Martell}, {Smith}, \&
  {Briley}}]{mar08b}
---. 2008{\natexlab{b}}, \pasp, 120, 7

\bibitem[{{Martell} {et~al.}(2008{\natexlab{c}}){Martell}, {Smith}, \&
  {Briley}}]{mar08c}
---. 2008{\natexlab{c}}, \aj, 136, 2522

\bibitem[{{Masseron} {et~al.}(2014){Masseron}, {Plez}, {Van Eck}, {Colin},
  {Daoutidis}, {Godefroid}, {Coheur}, {Bernath}, {Jorissen}, \&
  {Christlieb}}]{mas14}
{Masseron}, T., {Plez}, B., {Van Eck}, S., {et~al.} 2014, \aap, 571, A47

\bibitem[{{Mateo} {et~al.}(1990){Mateo}, {Harris}, {Nemec}, \&
  {Olszewski}}]{mat90}
{Mateo}, M., {Harris}, H.~C., {Nemec}, J., \& {Olszewski}, E.~W. 1990, \aj,
  100, 469

\bibitem[{{Matsunaga}(2006)}]{mat06}
{Matsunaga}, N. 2006, PhD thesis, University of Tokyo.

\bibitem[{{McClure} {et~al.}(1987){McClure}, {Hesser}, {Stetson}, {Vandenberg},
  \& {Bell}}]{mcc87}
{McClure}, R.~D., {Hesser}, J.~E., {Stetson}, P.~B., {Vandenberg}, D.~A., \&
  {Bell}, R.~A. 1987, \aj, 93, 1144

\bibitem[{{McClure} \& {Woodsworth}(1990)}]{mcc90}
{McClure}, R.~D., \& {Woodsworth}, A.~W. 1990, \apj, 352, 709

\bibitem[{{Mighell} \& {Burke}(1999)}]{mig99}
{Mighell}, K.~J., \& {Burke}, C.~J. 1999, \aj, 118, 366

\bibitem[{{Mouhcine}(2002)}]{mou02}
{Mouhcine}, M. 2002, \aap, 394, 125

\bibitem[{{Mouhcine} \& {Lan{\c c}on}(2003)}]{mou03}
{Mouhcine}, M., \& {Lan{\c c}on}, A. 2003, \mnras, 338, 572

\bibitem[{{Mould} \& {Aaronson}(1979)}]{mou79}
{Mould}, J., \& {Aaronson}, M. 1979, \apj, 232, 421

\bibitem[{{Mucciarelli} {et~al.}(2012){Mucciarelli}, {Bellazzini}, {Ibata},
  {Merle}, {Chapman}, {Dalessandro}, \& {Sollima}}]{muc12}
{Mucciarelli}, A., {Bellazzini}, M., {Ibata}, R., {et~al.} 2012, \mnras, 426,
  2889

\bibitem[{{Newman} {et~al.}(2013){Newman}, {Cooper}, {Davis}, {Faber}, {Coil},
  {Guhathakurta}, {Koo}, {Phillips}, {Conroy}, {Dutton}, {Finkbeiner}, {Gerke},
  {Rosario}, {Weiner}, {Willmer}, {Yan}, {Harker}, {Kassin}, {Konidaris},
  {Lai}, {Madgwick}, {Noeske}, {Wirth}, {Connolly}, {Kaiser}, {Kirby},
  {Lemaux}, {Lin}, {Lotz}, {Luppino}, {Marinoni}, {Matthews}, {Metevier}, \&
  {Schiavon}}]{new13}
{Newman}, J.~A., {Cooper}, M.~C., {Davis}, M., {et~al.} 2013, \apjs, 208, 5

\bibitem[{{Nomoto} {et~al.}(2006){Nomoto}, {Tominaga}, {Umeda}, {Kobayashi}, \&
  {Maeda}}]{nom06}
{Nomoto}, K., {Tominaga}, N., {Umeda}, H., {Kobayashi}, C., \& {Maeda}, K.
  2006, Nuclear Physics A, 777, 424

\bibitem[{{Norris} {et~al.}(2010){Norris}, {Wyse}, {Gilmore}, {Yong}, {Frebel},
  {Wilkinson}, {Belokurov}, \& {Zucker}}]{nor10}
{Norris}, J.~E., {Wyse}, R.~F.~G., {Gilmore}, G., {et~al.} 2010, \apj, 723,
  1632

\bibitem[{{Orban} {et~al.}(2008){Orban}, {Gnedin}, {Weisz}, {Skillman},
  {Dolphin}, \& {Holtzman}}]{orb08}
{Orban}, C., {Gnedin}, O.~Y., {Weisz}, D.~R., {et~al.} 2008, \apj, 686, 1030

\bibitem[{{Peterson} {et~al.}(1993){Peterson}, {Dalle Ore}, \&
  {Kurucz}}]{pet93}
{Peterson}, R.~C., {Dalle Ore}, C.~M., \& {Kurucz}, R.~L. 1993, \apj, 404, 333

\bibitem[{{Pietrzy{\'n}ski} {et~al.}(2008){Pietrzy{\'n}ski}, {Gieren},
  {Szewczyk}, {Walker}, {Rizzi}, {Bresolin}, {Kudritzki}, {Nalewajko}, {Storm},
  {Dall'Ora}, \& {Ivanov}}]{pie08}
{Pietrzy{\'n}ski}, G., {Gieren}, W., {Szewczyk}, O., {et~al.} 2008, \aj, 135,
  1993

\bibitem[{{Piskunov} {et~al.}(1995){Piskunov}, {Kupka}, {Ryabchikova}, {Weiss},
  \& {Jeffery}}]{pis95}
{Piskunov}, N.~E., {Kupka}, F., {Ryabchikova}, T.~A., {Weiss}, W.~W., \&
  {Jeffery}, C.~S. 1995, \aaps, 112, 525

\bibitem[{{Placco} {et~al.}(2014){Placco}, {Frebel}, {Beers}, \&
  {Stancliffe}}]{pla14}
{Placco}, V.~M., {Frebel}, A., {Beers}, T.~C., \& {Stancliffe}, R.~J. 2014,
  \apj, 797, 21

\bibitem[{{Revaz} \& {Jablonka}(2012)}]{rev12}
{Revaz}, Y., \& {Jablonka}, P. 2012, \aap, 538, A82

\bibitem[{{Revaz} {et~al.}(2009){Revaz}, {Jablonka}, {Sawala}, {Hill},
  {Letarte}, {Irwin}, {Battaglia}, {Helmi}, {Shetrone}, {Tolstoy}, \&
  {Venn}}]{rev09}
{Revaz}, Y., {Jablonka}, P., {Sawala}, T., {et~al.} 2009, \aap, 501, 189

\bibitem[{{Rizzi} {et~al.}(2007){Rizzi}, {Held}, {Saviane}, {Tully}, \&
  {Gullieuszik}}]{riz07}
{Rizzi}, L., {Held}, E.~V., {Saviane}, I., {Tully}, R.~B., \& {Gullieuszik}, M.
  2007, \mnras, 380, 1255

\bibitem[{{Robertson} {et~al.}(2005){Robertson}, {Bullock}, {Font}, {Johnston},
  \& {Hernquist}}]{rob05}
{Robertson}, B., {Bullock}, J.~S., {Font}, A.~S., {Johnston}, K.~V., \&
  {Hernquist}, L. 2005, \apj, 632, 872

\bibitem[{{Salpeter}(1955)}]{sal55}
{Salpeter}, E.~E. 1955, \apj, 121, 161

\bibitem[{{Sch{\"o}rck} {et~al.}(2009){Sch{\"o}rck}, {Christlieb}, {Cohen},
  {Beers}, {Shectman}, {Thompson}, {McWilliam}, {Bessell}, {Norris},
  {Mel{\'e}ndez}, {Ram{\'{\i}}rez}, {Haynes}, {Cass}, {Hartley}, {Russell},
  {Watson}, {Zickgraf}, {Behnke}, {Fechner}, {Fuhrmeister}, {Barklem},
  {Edvardsson}, {Frebel}, {Wisotzki}, \& {Reimers}}]{sch09}
{Sch{\"o}rck}, T., {Christlieb}, N., {Cohen}, J.~G., {et~al.} 2009, \aap, 507,
  817

\bibitem[{{Searle} \& {Zinn}(1978)}]{sea78}
{Searle}, L., \& {Zinn}, R. 1978, \apj, 225, 357

\bibitem[{{S{\'e}gall} {et~al.}(2007){S{\'e}gall}, {Ibata}, {Irwin}, {Martin},
  \& {Chapman}}]{seg07}
{S{\'e}gall}, M., {Ibata}, R.~A., {Irwin}, M.~J., {Martin}, N.~F., \&
  {Chapman}, S. 2007, \mnras, 375, 831

\bibitem[{{Sharina} {et~al.}(2012){Sharina}, {Aringer}, {Davoust}, {Kniazev},
  \& {Donzelli}}]{sha12}
{Sharina}, M., {Aringer}, B., {Davoust}, E., {Kniazev}, A.~Y., \& {Donzelli},
  C.~J. 2012, \mnras, 426, L31

\bibitem[{{Shetrone} {et~al.}(2003){Shetrone}, {Venn}, {Tolstoy}, {Primas},
  {Hill}, \& {Kaufer}}]{she03}
{Shetrone}, M., {Venn}, K.~A., {Tolstoy}, E., {et~al.} 2003, \aj, 125, 684

\bibitem[{{Shetrone} {et~al.}(1998{\natexlab{a}}){Shetrone}, {Bolte}, \&
  {Stetson}}]{she98a}
{Shetrone}, M.~D., {Bolte}, M., \& {Stetson}, P.~B. 1998{\natexlab{a}}, \aj,
  115, 1888

\bibitem[{{Shetrone} {et~al.}(1998{\natexlab{b}}){Shetrone}, {Briley}, \&
  {Brewer}}]{she98b}
{Shetrone}, M.~D., {Briley}, M., \& {Brewer}, J.~P. 1998{\natexlab{b}}, \aap,
  335, 919

\bibitem[{{Shetrone} {et~al.}(2001){Shetrone}, {C{\^o}t{\'e}}, \&
  {Sargent}}]{she01}
{Shetrone}, M.~D., {C{\^o}t{\'e}}, P., \& {Sargent}, W.~L.~W. 2001, \apj, 548,
  592

\bibitem[{{Shetrone} {et~al.}(2013){Shetrone}, {Smith}, {Stanford}, {Siegel},
  \& {Bond}}]{she13}
{Shetrone}, M.~D., {Smith}, G.~H., {Stanford}, L.~M., {Siegel}, M.~H., \&
  {Bond}, H.~E. 2013, \aj, 145, 123

\bibitem[{{Sills} {et~al.}(2013){Sills}, {Glebbeek}, {Chatterjee}, \&
  {Rasio}}]{sil13}
{Sills}, A., {Glebbeek}, E., {Chatterjee}, S., \& {Rasio}, F.~A. 2013, \apj,
  777, 105

\bibitem[{{Simon} \& {Geha}(2007)}]{sim07}
{Simon}, J.~D., \& {Geha}, M. 2007, \apj, 670, 313

\bibitem[{{Sk{\'u}lad{\'o}ttir} {et~al.}(2015){Sk{\'u}lad{\'o}ttir}, {Tolstoy},
  {Salvadori}, {Hill}, {Pettini}, {Shetrone}, \& {Starkenburg}}]{sku15}
{Sk{\'u}lad{\'o}ttir}, {\'A}., {Tolstoy}, E., {Salvadori}, S., {et~al.} 2015,
  \aap, 574, A129

\bibitem[{{Smith}(1984)}]{smi84}
{Smith}, G.~H. 1984, \aj, 89, 801

\bibitem[{{Smith} \& {Briley}(2005)}]{smi05}
{Smith}, G.~H., \& {Briley}, M.~M. 2005, \pasp, 117, 895

\bibitem[{{Smith} \& {Briley}(2006)}]{smi06}
---. 2006, \pasp, 118, 740

\bibitem[{{Smith} {et~al.}(1996){Smith}, {Shetrone}, {Bell}, {Churchill}, \&
  {Briley}}]{smi96}
{Smith}, G.~H., {Shetrone}, M.~D., {Bell}, R.~A., {Churchill}, C.~W., \&
  {Briley}, M.~M. 1996, \aj, 112, 1511

\bibitem[{{Sneden} {et~al.}(1992){Sneden}, {Kraft}, {Prosser}, \&
  {Langer}}]{sne92}
{Sneden}, C., {Kraft}, R.~P., {Prosser}, C.~F., \& {Langer}, G.~E. 1992, \aj,
  104, 2121

\bibitem[{{Sneden}(1973)}]{sne73}
{Sneden}, C.~A. 1973, PhD thesis, University of Texas Austin.

\bibitem[{{Sobeck} {et~al.}(2011){Sobeck}, {Kraft}, {Sneden}, {Preston},
  {Cowan}, {Smith}, {Thompson}, {Shectman}, \& {Burley}}]{sob11}
{Sobeck}, J.~S., {Kraft}, R.~P., {Sneden}, C., {et~al.} 2011, \aj, 141, 175

\bibitem[{{Spite} {et~al.}(2005){Spite}, {Cayrel}, {Plez}, {Hill}, {Spite},
  {Depagne}, {Fran{\c c}ois}, {Bonifacio}, {Barbuy}, {Beers}, {Andersen},
  {Molaro}, {Nordstr{\"o}m}, \& {Primas}}]{spi05}
{Spite}, M., {Cayrel}, R., {Plez}, B., {et~al.} 2005, \aap, 430, 655

\bibitem[{{Spite} {et~al.}(2006){Spite}, {Cayrel}, {Hill}, {Spite}, {Fran{\c
  c}ois}, {Plez}, {Bonifacio}, {Molaro}, {Depagne}, {Andersen}, {Barbuy},
  {Beers}, {Nordstr{\"o}m}, \& {Primas}}]{spi06}
{Spite}, M., {Cayrel}, R., {Hill}, V., {et~al.} 2006, \aap, 455, 291

\bibitem[{{Starkenburg} {et~al.}(2014){Starkenburg}, {Shetrone}, {McConnachie},
  \& {Venn}}]{sta14}
{Starkenburg}, E., {Shetrone}, M.~D., {McConnachie}, A.~W., \& {Venn}, K.~A.
  2014, \mnras, 441, 1217

\bibitem[{{Starkenburg} {et~al.}(2010){Starkenburg}, {Hill}, {Tolstoy},
  {Gonz{\'a}lez Hern{\'a}ndez}, {Irwin}, {Helmi}, {Battaglia}, {Jablonka},
  {Tafelmeyer}, {Shetrone}, {Venn}, \& {de Boer}}]{sta10}
{Starkenburg}, E., {Hill}, V., {Tolstoy}, E., {et~al.} 2010, \aap, 513, A34

\bibitem[{{Starkenburg} {et~al.}(2013){Starkenburg}, {Hill}, {Tolstoy},
  {Fran{\c c}ois}, {Irwin}, {Boschman}, {Venn}, {de Boer}, {Lemasle},
  {Jablonka}, {Battaglia}, {Groot}, \& {Kaper}}]{sta13}
---. 2013, \aap, 549, A88

\bibitem[{{Stetson}(2000)}]{ste00}
{Stetson}, P.~B. 2000, \pasp, 112, 925

\bibitem[{{Stetson} {et~al.}(1998){Stetson}, {Hesser}, \&
  {Smecker-Hane}}]{ste98}
{Stetson}, P.~B., {Hesser}, J.~E., \& {Smecker-Hane}, T.~A. 1998, \pasp, 110,
  533

\bibitem[{{Suntzeff}(1981)}]{sun81}
{Suntzeff}, N.~B. 1981, \apjs, 47, 1

\bibitem[{{Suntzeff} {et~al.}(1988){Suntzeff}, {Kraft}, \& {Kinman}}]{sun88}
{Suntzeff}, N.~B., {Kraft}, R.~P., \& {Kinman}, T.~D. 1988, \aj, 95, 91

\bibitem[{{Takeda} {et~al.}(2009){Takeda}, {Kaneko}, {Matsumoto}, {Oshino},
  {Ito}, \& {Shibuya}}]{tak09}
{Takeda}, Y., {Kaneko}, H., {Matsumoto}, N., {et~al.} 2009, \pasj, 61, 563

\bibitem[{{Tinsley}(1980)}]{tin80}
{Tinsley}, B.~M. 1980, \fcp, 5, 287

\bibitem[{{van Loon} {et~al.}(2007){van Loon}, {van Leeuwen}, {Smalley},
  {Smith}, {Lyons}, {McDonald}, \& {Boyer}}]{van07}
{van Loon}, J.~T., {van Leeuwen}, F., {Smalley}, B., {et~al.} 2007, \mnras,
  382, 1353

\bibitem[{{VandenBerg} {et~al.}(2012){VandenBerg}, {Bergbusch}, {Dotter},
  {Ferguson}, {Michaud}, {Richer}, \& {Proffitt}}]{van12}
{VandenBerg}, D.~A., {Bergbusch}, P.~A., {Dotter}, A., {et~al.} 2012, \apj,
  755, 15

\bibitem[{{Venn} {et~al.}(2004){Venn}, {Irwin}, {Shetrone}, {Tout}, {Hill}, \&
  {Tolstoy}}]{ven04}
{Venn}, K.~A., {Irwin}, M., {Shetrone}, M.~D., {et~al.} 2004, \aj, 128, 1177

\bibitem[{{Venn} {et~al.}(2012){Venn}, {Shetrone}, {Irwin}, {Hill}, {Jablonka},
  {Tolstoy}, {Lemasle}, {Divell}, {Starkenburg}, {Letarte}, {Baldner},
  {Battaglia}, {Helmi}, {Kaufer}, \& {Primas}}]{ven12}
{Venn}, K.~A., {Shetrone}, M.~D., {Irwin}, M.~J., {et~al.} 2012, \apj, 751, 102

\bibitem[{{Villanova} {et~al.}(2007){Villanova}, {Piotto}, {King}, {Anderson},
  {Bedin}, {Gratton}, {Cassisi}, {Momany}, {Bellini}, {Cool}, {Recio-Blanco},
  \& {Renzini}}]{vil07}
{Villanova}, S., {Piotto}, G., {King}, I.~R., {et~al.} 2007, \apj, 663, 296

\bibitem[{{Walker}(1994)}]{wal94}
{Walker}, A.~R. 1994, \aj, 108, 555

\bibitem[{{Wallerstein} \& {Knapp}(1998)}]{wal98}
{Wallerstein}, G., \& {Knapp}, G.~R. 1998, \araa, 36, 369

\bibitem[{{Westfall} {et~al.}(2006){Westfall}, {Majewski}, {Ostheimer},
  {Frinchaboy}, {Kunkel}, {Patterson}, \& {Link}}]{wes06}
{Westfall}, K.~B., {Majewski}, S.~R., {Ostheimer}, J.~C., {et~al.} 2006, \aj,
  131, 375

\bibitem[{{Winnick}(2003)}]{win03}
{Winnick}, R.~A. 2003, PhD thesis, Yale University

\bibitem[{{Yanny} {et~al.}(2009){Yanny}, {Rockosi}, {Newberg}, {Knapp},
  {Adelman-McCarthy}, {Alcorn}, {Allam}, {Allende Prieto}, {An}, {Anderson},
  {Anderson}, {Bailer-Jones}, {Bastian}, {Beers}, {Bell}, {Belokurov},
  {Bizyaev}, {Blythe}, {Bochanski}, {Boroski}, {Brinchmann}, {Brinkmann},
  {Brewington}, {Carey}, {Cudworth}, {Evans}, {Evans}, {Gates}, {G{\"a}nsicke},
  {Gillespie}, {Gilmore}, {Nebot Gomez-Moran}, {Grebel}, {Greenwell}, {Gunn},
  {Jordan}, {Jordan}, {Harding}, {Harris}, {Hendry}, {Holder}, {Ivans},
  {Ivezi{\v c}}, {Jester}, {Johnson}, {Kent}, {Kleinman}, {Kniazev},
  {Krzesinski}, {Kron}, {Kuropatkin}, {Lebedeva}, {Lee}, {French Leger},
  {L{\'e}pine}, {Levine}, {Lin}, {Long}, {Loomis}, {Lupton}, {Malanushenko},
  {Malanushenko}, {Margon}, {Martinez-Delgado}, {McGehee}, {Monet}, {Morrison},
  {Munn}, {Neilsen}, {Nitta}, {Norris}, {Oravetz}, {Owen}, {Padmanabhan},
  {Pan}, {Peterson}, {Pier}, {Platson}, {Re Fiorentin}, {Richards}, {Rix},
  {Schlegel}, {Schneider}, {Schreiber}, {Schwope}, {Sibley}, {Simmons},
  {Snedden}, {Allyn Smith}, {Stark}, {Stauffer}, {Steinmetz}, {Stoughton},
  {SubbaRao}, {Szalay}, {Szkody}, {Thakar}, {Sivarani}, {Tucker}, {Uomoto},
  {Vanden Berk}, {Vidrih}, {Wadadekar}, {Watters}, {Wilhelm}, {Wyse}, {Yarger},
  \& {Zucker}}]{yan09}
{Yanny}, B., {Rockosi}, C., {Newberg}, H.~J., {et~al.} 2009, \aj, 137, 4377

\bibitem[{{Yong} {et~al.}(2013){Yong}, {Norris}, {Bessell}, {Christlieb},
  {Asplund}, {Beers}, {Barklem}, {Frebel}, \& {Ryan}}]{yon13}
{Yong}, D., {Norris}, J.~E., {Bessell}, M.~S., {et~al.} 2013, \apj, 762, 27

\bibitem[{{Zinn}(1981)}]{zin81}
{Zinn}, R. 1981, \apj, 251, 52

\end{thebibliography}
\bibliographystyle{apj}

\clearpage
\renewcommand{\thetable}{\arabic{table}}
\setcounter{table}{3}
\begin{turnpage}

\begin{deluxetable}{llcccccccccccc}
\tablewidth{0pt}
\tablecolumns{14}
\tablecaption{Carbon Abundances\label{tab:cfe}}
\tablehead{\colhead{System} & \colhead{Name} & \colhead{RA} & \colhead{Dec} & \colhead{$\log L$} & \colhead{S/N} & \colhead{$T_{\rm eff}$} & \colhead{$\log g$} & \colhead{[Fe/H]} & \colhead{[$\alpha$/Fe]} & \colhead{[C/Fe]} & \colhead{${\rm [C/Fe]}_{\rm corr}$}\tablenotemark{a} & \colhead{$S_2({\rm CH})$} & \colhead{C star?} \\
                            &                & (J2000)      & (J2000)       & ($L_{\sun}$)       & (\AA$^{-1}$)  & (K)                     & (cm~s$^{-2}$)      & (dex)            & (dex)                   & (dex)            & (dex)                                                & (mag)                     &                  }
\startdata
Ursa Minor & Bel60031      & 15 07 38.73 & $+67$ 13 56.4 & 2.94 &         &      &      &                  &                  &                   &          &                 & C \\
Ursa Minor & Bel60021      & 15 07 50.88 & $+67$ 11 12.9 & 2.63 & \phn 99 & 4495 & 1.30 & $-2.43 \pm 0.11$ & $+0.08 \pm 0.15$ & $ -0.65 \pm 0.16$ & $ -0.01$ &     \nodata     &   \\
Ursa Minor & Bel60469      & 15 07 55.52 & $+67$ 11 30.2 & 1.74 & \phn 32 & 4772 & 2.33 & $-2.78 \pm 0.25$ &     \nodata      & $<+0.44         $ & $<+0.45$ & $1.75 \pm 0.33$ &   \\
Ursa Minor & Bel60230      & 15 07 59.77 & $+67$ 15 28.8 & 2.28 &         &      &      &                  &                  &                   &          &                 & C \\
Ursa Minor & Bel60123      & 15 07 59.77 & $+67$ 11 41.4 & 2.10 & \phn 56 & 4826 & 1.96 & $-2.36 \pm 0.13$ & $+0.35 \pm 0.19$ & $<+0.06         $ & $<+0.13$ & $1.60 \pm 0.11$ &   \\
Ursa Minor & Bel70584      & 15 08 08.20 & $+67$ 06 43.8 & 1.62 & \phn 24 & 4899 & 2.49 & $-2.30 \pm 0.21$ & $-0.19 \pm 0.68$ & $<+0.55         $ & $<+0.56$ &     \nodata     &   \\
Ursa Minor & Bel70182      & 15 08 08.29 & $+67$ 04 41.0 & 2.00 & \phn 25 & 4803 & 2.06 & $-2.02 \pm 0.16$ & $-0.17 \pm 0.46$ & $<+0.90         $ & $<+0.94$ &     \nodata     &   \\
Ursa Minor & Bel60017      & 15 08 09.06 & $+67$ 09 21.6 & 2.63 & \phn 70 & 4526 & 1.33 & $-2.07 \pm 0.11$ & $+0.17 \pm 0.13$ & $ +0.31 \pm 0.27$ & $ +0.73$ & $2.13 \pm 0.05$ &   \\
Ursa Minor & Bel50017      & 15 08 10.51 & $+67$ 17 07.1 & 2.60 &     111 & 4529 & 1.35 & $-2.18 \pm 0.11$ & $+0.04 \pm 0.08$ & $ -0.70 \pm 0.21$ & $ -0.10$ & $1.59 \pm 0.05$ &   \\
Ursa Minor & Bel60819      & 15 08 12.07 & $+67$ 10 15.5 & 1.27 & \phn 13 & 5198 & 2.93 & $-1.85 \pm 0.33$ &     \nodata      & $<-0.26         $ & $<-0.25$ & $1.54 \pm 0.17$ &   \\
\nodata & \nodata & \nodata & \nodata & \nodata & \nodata & \nodata & \nodata & \nodata & \nodata & \nodata & \nodata & \nodata & \nodata \\
\enddata
\tablerefs{Photometry comes from the following sources: NGC~2419 and M15 \protect\citep{ste00}, M68 \protect\citep{alc90,wal94,ste00}, Sculptor \protect\citep{wes06}, Fornax \protect\citep{ste98}, Ursa Minor \protect\citep{bel02}, Draco \protect\citep{seg07}.}
\tablecomments{(This table is available in its entirety in a machine-readable form in the online journal. A portion is shown here for guidance regarding its form and content.  The online version of the table has additional columns for photometric magnitudes.)}
\tablenotetext{a}{[C/Fe] corrected for stellar evolutionary state according to \protect\citet{pla14}.}
\end{deluxetable}

\clearpage
\end{turnpage}

\end{document}